\documentclass[11pt,a4paper]{article}
\pdfoutput=1
\usepackage{jcappub}
\usepackage{booktabs}
\usepackage[english]{babel}
\usepackage{amsmath,amssymb,amsbsy,amstext, amsthm, simplewick}
\usepackage{graphicx}
\usepackage{float}
\usepackage{amsfonts}
\usepackage{amssymb}
\usepackage{braket} 
\usepackage{upgreek}
 \usepackage{exscale,relsize}
 \usepackage[makeroom]{cancel}
\usepackage{soul}
\usepackage[hang,flushmargin]{footmisc}

\allowdisplaybreaks[1]

\usepackage[top=2.5cm, bottom=3cm, outer=2cm, inner=2cm]{geometry}

\newcommand{\beq}{\begin{equation}}
\newcommand{\eeq}{\end{equation}}
\newcommand{\beqa}{\begin{align}}
\newcommand{\eeqa}{\end{align}}

\newcommand{\calO}{\mathcal{O}}

\newcommand{\calF}{\mathcal{F}}

\newcommand{\calP}{\mathcal{P}}
\newcommand{\calN}{\mathcal{N}}
\newcommand{\calI}{\mathcal{I}}
\newcommand{\p}{\partial}

\newcommand{\scpt}{\scriptscriptstyle}
\newcommand{\Kahler}{K\"{a}hler }
\newcommand{\Mp}{M_{\scriptscriptstyle P\ell}}
\newcommand{\phiNe}{\phi_{\scpt\calN_e}}
\newcommand{\fN}{f_{\scpt N}}
\newcommand{\vev}[1]{ \left\langle {#1} \right\rangle }

\makeatletter
\def\blfootnote{\xdef\@thefnmark{}\@footnotetext}
\makeatother

\begin{document}

\begin{titlepage}

\newgeometry{top=2.5cm, bottom=1cm, outer=2cm, inner=2cm}

\setcounter{page}{1} \baselineskip=15.5pt \thispagestyle{empty}

\bigskip\

\vspace{1cm}
\begin{center}

{\fontsize{20}{24}\selectfont  \sffamily \bfseries  New Inflation in the Landscape and \vspace{10pt} \\  Typicality of the Observed Cosmic Perturbation}

\end{center}

\vspace{0.2cm}
\begin{center}
{\fontsize{14}{30}\selectfont Chien-I Chiang$^{\bigstar\diamondsuit}$ and Keisuke Harigaya$^{\bigstar\diamondsuit\clubsuit}$} 
\end{center}

\begin{center}

\vskip 8pt
\textsl{$^\bigstar$Berkeley Center for Theoretical Physics, Department of Physics, University of California, Berkeley, CA 94720, USA} 

\vskip 6pt

\textsl{$^\diamondsuit$Theoretical Physics Group, Lawrence Berkeley National Laboratory,
Berkeley, California, USA} 

\vskip 6pt

\textsl{$^\clubsuit$School of Natural Sciences, Institute for Advanced Study, Princeton, NJ 08540, USA}

\vskip 7pt

\end{center}

\vspace{1.2cm}
\noindent \rule{\textwidth}{1pt} \vspace{0.3cm}
\noindent {\sffamily \bfseries Abstract} \\[0.1cm]
We investigate if the observed small and nearly scale-invariant primordial cosmic perturbation, i.e.\  the perturbation amplitude $P_\zeta\sim10^{-9}$ and the spectral index $n_s \simeq 0.965$, is typical in the landscape of vacua after imposing anthropic selections on them. We consider the situation where the universe begins from a metastable vacuum driving a precedent inflation, a curvature-dominated open universe is created by tunneling, and the curvature energy is inflated away by new inflation. We argue that the initial inflaton field value is homogeneous but typically non-zero because of the quantum fluctuation of long wavelength modes created during the precedent inflation, and only the universe which accidentally has a small inflaton field value is anthropically selected. We show that this bias, together with certain distributions of inflation model parameters that are physically well-motivated, makes the observed small and nearly scale-invariant spectrum typical.
 
\vskip 10pt
\noindent \rule{\textwidth}{1pt}
\vskip 10pt

\vspace{0.6cm}

\blfootnote{ \href{mailto:chienichiang@berkeley.edu}{chienichiang@berkeley.edu} }
\blfootnote{ \href{mailto:keisukeharigaya@ias.edu }{keisukeharigaya@ias.edu } }

 \end{titlepage}

\newgeometry{top=2.5cm, bottom=3cm, outer=2cm, inner=2cm}

\noindent \rule{\textwidth}{1pt}

\tableofcontents

\vspace{0.5cm}

\noindent \rule{\textwidth}{1pt}

\section{Introduction}

Inflationary paradigm not only solves the horizon and flatness problem~\cite{Guth:1980zm} (see also~\cite{Kazanas:1980tx}), but also elegantly explains the nearly scale-invariant and Gaussian cosmic perturbation imprinted in the cosmic microwave background (CMB) and the large scale structure of the universe~\cite{Mukhanov:1981xt,Hawking:1982cz,Starobinsky:1982ee,Guth:1982ec,Bardeen:1983qw}, given that inflation is driven by a scalar field with a very flat potential~\cite{Linde:1981mu,Albrecht:1982wi} (see also~\cite{Starobinsky:1980te}). However, despite the phenomenological success of the generic paradigm, the underlying physical origin of cosmic inflation is still an open problem.

We investigate the inflation paradigm in the view point of the string landscape (see~\cite{Susskind:2003kw} for a review). The string theory predicts that there are numerous vacua, and each vacuum yields an effective field theory with a different set of fields and parameters. An example leading to various cosmological constants is given in~\cite{Bousso:2000xa}. The landscape of vacua supports the notion of the anthropic principle. The parameters of the nature which we observe is not necessarily explained by the dynamics of the theory, but may be chosen so that the human civilization can exist. There would be multiple vacua on which we can live. We can calculate the distribution function of the parameters sampled from those habitable vacua weighted by the number of observers in the vacua. The parameter we observed would be around the most plausible one (the principle of mediocrity~\cite{Vilenkin:1994ua}). This notion succeeded in predicting a rough value of the cosmological constant~\cite{Weinberg:1987dv}.

In the landscape the expected inflationary dynamics is the following~\cite{Freivogel:2005vv,Guth:2013sya}. The universe would be initially inhomogeneous, with length/energy scales set by the fundamental scale. A scalar field resides in a meta-stable vacuum and the potential energy eventually dominates the universe, driving a \textit{precedent inflation} which erases the inhomogeneity. The scalar field tunnels toward the vacuum with a small potential energy, and the universe becomes open and curvature dominated~\cite{Coleman:1980aw,Gott:1982zf}. For habitability, inflation with a sufficient number of e-foldings must occur afterward, since otherwise the galaxy formation is prevented~\cite{Linde:1995rv,Vilenkin:1996ar}. Then the flatness of the inflaton potential is not necessarily the one to be explained by the property of the theory, but may be as a result of the anthropic selection. Still, we should ask if the small, $P_\zeta \sim 10^{-9}$, and nearly scale-invariant, $n_s \sim0.96$, cosmic perturbation \cite{Aghanim:2018eyx} is a plausible one.
We investigate this question by considering the inflationary dynamics as well as the post-inflationary evolution of the universe.

Anthropic arguments from the post-inflationary evolution alone do not seem strong enough to enforce the amplitude of primordial perturbation power spectrum $P_\zeta \sim 10^{-9}$.
A larger energy density from cosmological constant $\rho_{\scpt \Lambda}$ requires a larger primordial perturbation so that structure can be formed in our universe. In particular, the density contrast at the time of matter-dark energy equality needs to be larger than a certain threshold to allow structure formation~\cite{Weinberg:1987dv,Martel:1997vi,Barrow:1993}
\beq
\left(\frac{\delta \rho}{\rho} \right)_{\rm M-DE \,eq} 
= \left(\frac{\delta \rho}{\rho} \right)_{\rm R-M \, eq} \left(\frac{a_{\scpt \rm M-DE \, eq}}{a_{\scpt \rm R-M \, eq}} \right) \propto  \left( \frac{\delta \rho}{\rho} \right)_* \left(\frac{\Omega^{(0)}_M}{\Omega^{(0)}_\Lambda} \right)^{1/3} > \left(\frac{\delta \rho}{\rho} \right)_{min}.
\eeq
Here the subscripts $\rm R-M\,eq$, $\rm M-DE \, eq$, and $*$ denote the time of radiation-matter equality, matter-dark energy equality, and the time of horizon re-entrance respectively. We approximated the density contrast $\delta \rho/\rho$ at the time of radiation-matter equality by that at horizon re-entrance because the density contrast only evolves logarithmically during radiation dominant era. With $(\delta \rho/\rho)^2 \sim P_\zeta$, this means that for a given $P_\zeta$, the maximum energy density the cosmological constant can have is then%
\footnote{There are other criteria proposed for the anthropic conditions for the dark energy density (see, e.g.,~\cite{Barrow:1988yia,Bousso:2010vi}), which can lead to different powers than $3/2$. In this paper we consider the original criterion in~\cite{Weinberg:1987dv,Martel:1997vi}.}
\beq
\rho_\Lambda^{max} \propto P_\zeta^{3/2}.
\eeq
Assuming the energy density of the cosmological constant follows a uniform probability distribution
\beq
\int^{\rho_\Lambda^{max}}_0 d\rho_{\scpt \Lambda},
\eeq
this translates to a contribution to the probability distribution of $P_\zeta$ of the form $P_\zeta^{3/2}$ which biases toward large $P_\zeta$.

A universe with a very large $P_\zeta$ may be anthropically disfavored by the property of the galaxy~\cite{Tegmark:1997in,Tegmark:2005dy}. If $(\delta \rho / \rho)$ is too large, 
the galaxy would be too dense such that the time scale of orbital disruption and close encounter with nearby planets is too short.
Although it is not clear what kind of encounter kills the earth-like planet, and the corresponding bound on $\delta \rho / \rho$ is uncertain, we adopt the bound of $(\delta \rho / \rho )<\calO(10^{-4})$~\cite{Tegmark:1997in,Tegmark:2005dy}. Although the typical value of $\delta \rho / \rho$ is $P_\zeta^{1/2}$, even if $P_\zeta>\calO(10^{-8})$, it is still possible that we live in a part of the universe with a small energy contrast. Assuming the probability distribution to populate at the region with a density contrast $\delta \equiv \delta \rho/\rho$ in a universe with a primordial perturbation amplitude $P_\zeta$ is Gaussian, the probability to be in the habitable region is%
\footnote{The property of the galaxy may depend on $P_\zeta$. For example, for larger $P_\zeta$ the formation of proto-galaxies occurs earlier, which will change the initial metallicity of the galaxy. We do not consider this effect in this paper.}
\beq
\int^{10^{-4}}_0 d\delta \, \frac{1}{\sqrt{2\pi P_\zeta}} e^{-\frac{\delta^2}{2P_\zeta}}
\simeq \int^{10^{-4}}_0 d\delta \, \frac{1}{\sqrt{2\pi P_\zeta}} \propto P_\zeta^{-1/2}.
\eeq

\begin{figure}[tb]
\begin{center}
\includegraphics[width=0.5\textwidth]{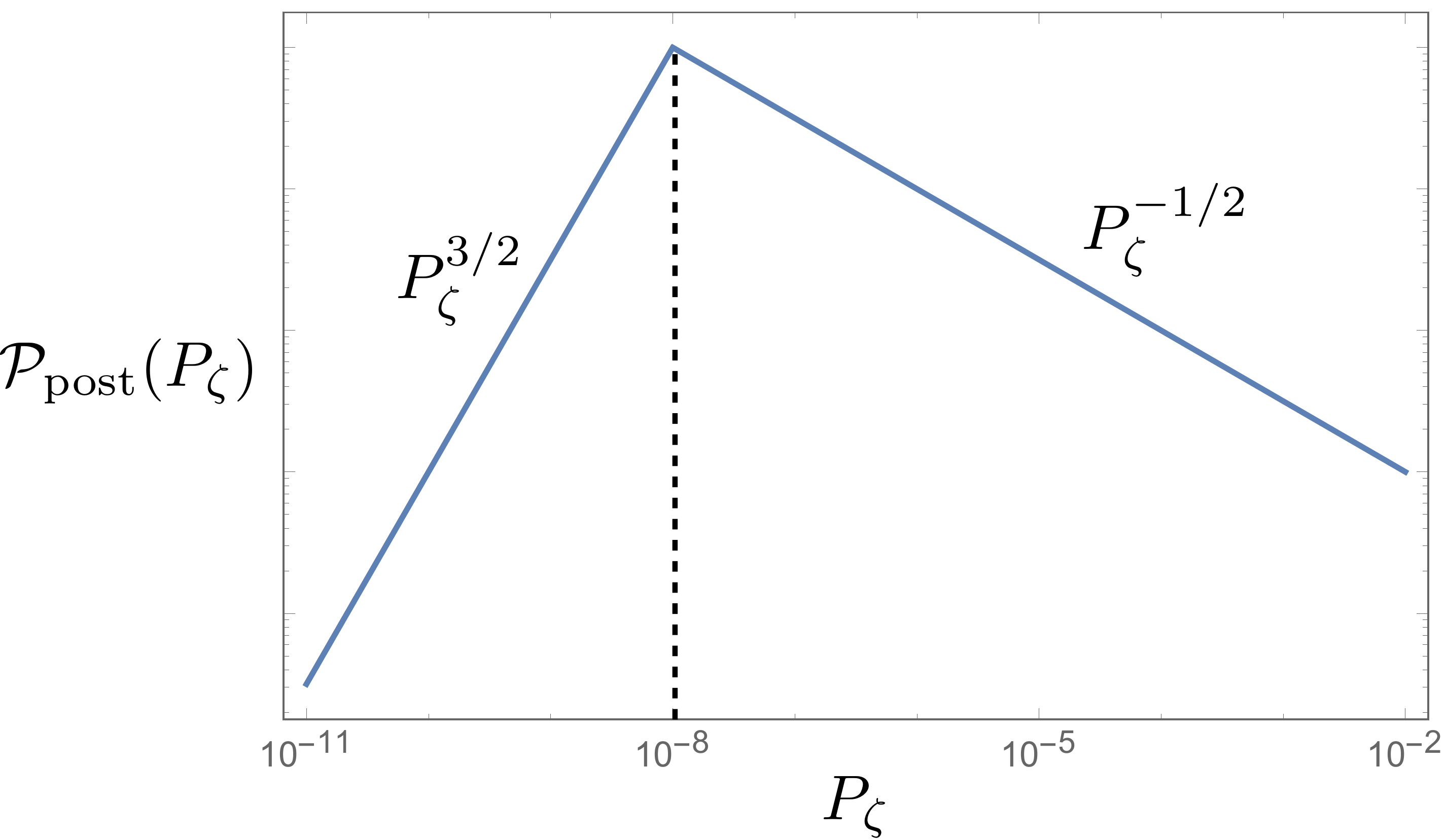}
\caption{The schematic summary of the probability distribution $\calP_{\rm post}(P_\zeta)$ of having a primordial perturbation amplitude $P_\zeta$ from anthropic consideration of post-inflationary evolution.}
\label{fig:PPzetaPost}
\end{center}
\end{figure}

In Figure \ref{fig:PPzetaPost} we schematically summarize the probability distribution $\calP_{\rm post}(P_\zeta)$ of having a universe with primordial perturbation amplitude $P_\zeta$ based on the anthropic consideration of post-inflationary evolution we discussed. We see that by just considering the post-inflationary evolution, the observed value $P_\zeta \sim 10^{-9}$ already require a fine-tuning of about a few percent.  In addition, to obtain the full probability of having a $P_\zeta$, one also needs to consider the probability stemming from inflation dynamics. In doing so, as most of the realistic measures suggest~\cite{Linde:1993xx,Bousso:2006ev,DeSimone:2008bq,Bousso:2008hz,Nomura:2011dt,Garriga:2012bc}, we do not weight the increase of the volume due to inflation. If the inflation scale $V$ is simply given by a mass parameter, it is biased toward the fundamental scale. Then $P_\zeta = \frac{V}{24\pi^2 \epsilon}$ is also biased toward larger values. 
For example, in the appendix, we consider a generic small field inflation model with $Z_2$ symmetry and find that the probability from anthropic consideration on inflation alone strongly bias toward large $P_\zeta$ with $P_{\rm inf}(P_\zeta) \propto P_\zeta^{19/4}$. In this type of model the small primordial perturbation is highly implausible. An inflation model with $P_\zeta$ not biased toward large values is required.

A spectral index close to unity is apparently challenging. The spectral index $n_s$ is given by 
\beq
n_s = 1+ 2\eta -6\epsilon,~\eta \equiv \frac{V''}{V},~\epsilon \equiv \frac{1}{2}\left(\frac{V'}{V}\right)^2.
\eeq
Having a spectral index $n_s \sim 0.96$ requires $\eta \sim 0.02 \ll 1$. In order to explain the nearly-scale-invariant spectrum, we essentially need to solve the $\eta$-problem~\cite{Ovrut:1983my,Holman:1984yj,Goncharov:1983mw,Coughlan:1984yk,Copeland:1994vg}. It is not obvious if the requirement of the large enough number of e-foldings can ensure such small $\eta$ parameter.

Ref.~\cite{Tegmark:2004qd} investigates the distributions of $P_\zeta$ and $n_s$, assuming that the inflaton potential obeys a Gaussian distribution, and find that the observed values are highly implausible unless the inflaton field value is as large as the Planck scale. Ref.~\cite{Masoumi:2016eag} investigates the inflection point inflation also assuming the Gaussian distribution.
In this set up the number of the e-folding tends to be larger for a small and positive $\eta$ parameter. After imposing the anthropic requirement, the spectrum tends to be blue, but the probability of $n_s < 0.97$ is found to be about $0.2$, which is reasonably high. However, the distribution of $P_\zeta$ is not discussed.

In this paper we investigate the distributions of $P_\zeta$ and $n_s$ for a new inflation model  where the inflaton is trapped around the origin during the precedent inflation by a Hubble induced mass. Although the field value of the inflaton is homogeneous inside the horizon because of the damping during the precedent inflation, quantum fluctuation of long-wave length modes is produced, which effectively works as a homogeneous but non-zero initial condition of inflation, unless the Hubble induced mass is much larger than the Hubble scale to suppress the quantum fluctuation. We show that this probabilistic nature of inflaton initial condition is an important key to understand $n_s$ close to unity.
We focus on a supersymmetric model. As we will see, the smallness of $P_\zeta$ is then also explained, as some of the parameter of the theory can be biased toward small values or logarithmically distributed in supersymmetric theories.
Our results are summarized in Figure \ref{fig:PkSpace}, where we show the probability distribution $P_\zeta \calP_{\rm net}(P_\zeta, k)$ of $P_\zeta$ and $k \simeq -\eta$, taking into account of both inflationary and post-inflationary dynamics. In the contour plot we can see that the probability distribution is biased toward smaller $k$ value and hence the $\eta$-problem is solved. In addition, with an anthropic bound on the density contrast $\delta \rho/\rho < \calO(10^{-4})$, the observed universe with $P_\zeta \sim 10^{-9}$ and $n_s \simeq 0.965$, which is marked by the blue star in Figure \ref{fig:PkSpace}, is actually a typical one.

This paper is organized as follows. In the next section
We first elaborate on the necessity of including the probability distribution of inflaton initial condition in generic new inflation models.
 We then consider a supersymmetric model and parametrize the probability distributions of the couplings. We find
 that, for a certain distribution of the couplings of the model, the observed small ($P_\zeta \sim 10^{-9}$) and scale-invariant $(n_s \simeq 0.96)$ curvature perturbation is probabilistically favorable.
We then discuss and summarize our results in Sec.\ref{Sec:Discussion}. In the appendix we show our study on the general new-inflation-type model with $Z_2$ symmetry, where observed spectral index $n_s$ is probabilistically favored but the smallness of primordial perturbation cannot be explained.

\section{New Inflation in the Landscape after Quantum Tunneling}\label{Sec:LandscapeInflation}

\begin{figure}[tb]
\begin{center}
\includegraphics[width=0.9\textwidth]{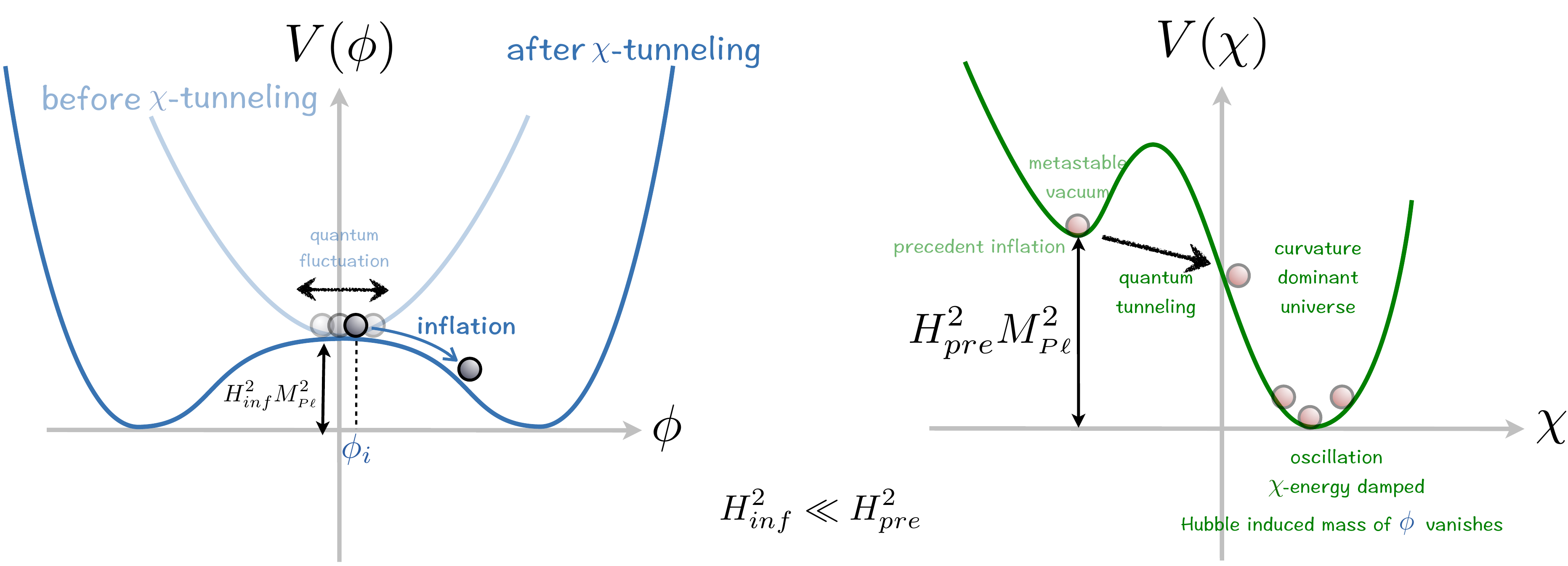}
\caption{Schematic figure of the proposed scenario.}
\label{fig:SUSYscenario}
\end{center}
\end{figure}

In this section we consider new-inflation-type models in the landscape. We assume that the last inflation which explains the flatness of the universe and the observed cosmic perturbation is a new-inflation-type model with an inflaton $\phi$. In the theory with multiple vacua, it is expected that a singlet scalar field $\chi$ stays at its metastable vacuum and drives a precedent inflationary expansion, leading to a homogeneous universe. After the quantum tunneling of the singlet scalar field, the universe becomes an infinite open curvature dominated Friedmann-Robertson-Walker (FRW) universe while the scalar field rolls down to a local minimum with a small potential energy. The universe is eventually dominated by the potential energy of the inflaton (Figure~\ref{fig:SUSYscenario}). One may naively expect that a coupling between the $\chi$ field and the inflaton (leading to so-called the Hubble induced mass) can trap the inflaton to the origin and the initial inflaton field value $\phi_i$ is automatically small enough to initiate the last inflation. This is generically not true. As we will see, after the tunneling the Hubble induced mass of the inflaton is not effective. Therefore the inflaton fluctuation mode that just exited the horizon before quantum tunneling may survive. Although the inflaton field value is homogeneous inside the horizon, the field value must be fine-tuned for the last inflation to occur and last long enough. We investigate the impact of this observation by computing the distribution function of the curvature perturbation $P_\zeta$ and the spectral index $n_s$ (equivalently the $\eta$ parameter) after requiring enough number of e-folds $\calN^{tot}_e$ during the last inflation.
We find that $P_\zeta$ as well as $\eta$ may be biased toward small values, explaining the observed very small ($P_\zeta \sim 10^{-9}$) and scale-invariant $(n_s \simeq 0.96)$ curvature perturbation.

\subsection{Hubble Induced Mass and the Initial Condition after Tunneling}\label{subsec:InitialCondition}
\begin{figure}[tb]
\begin{center}
\includegraphics[width=0.5\textwidth]{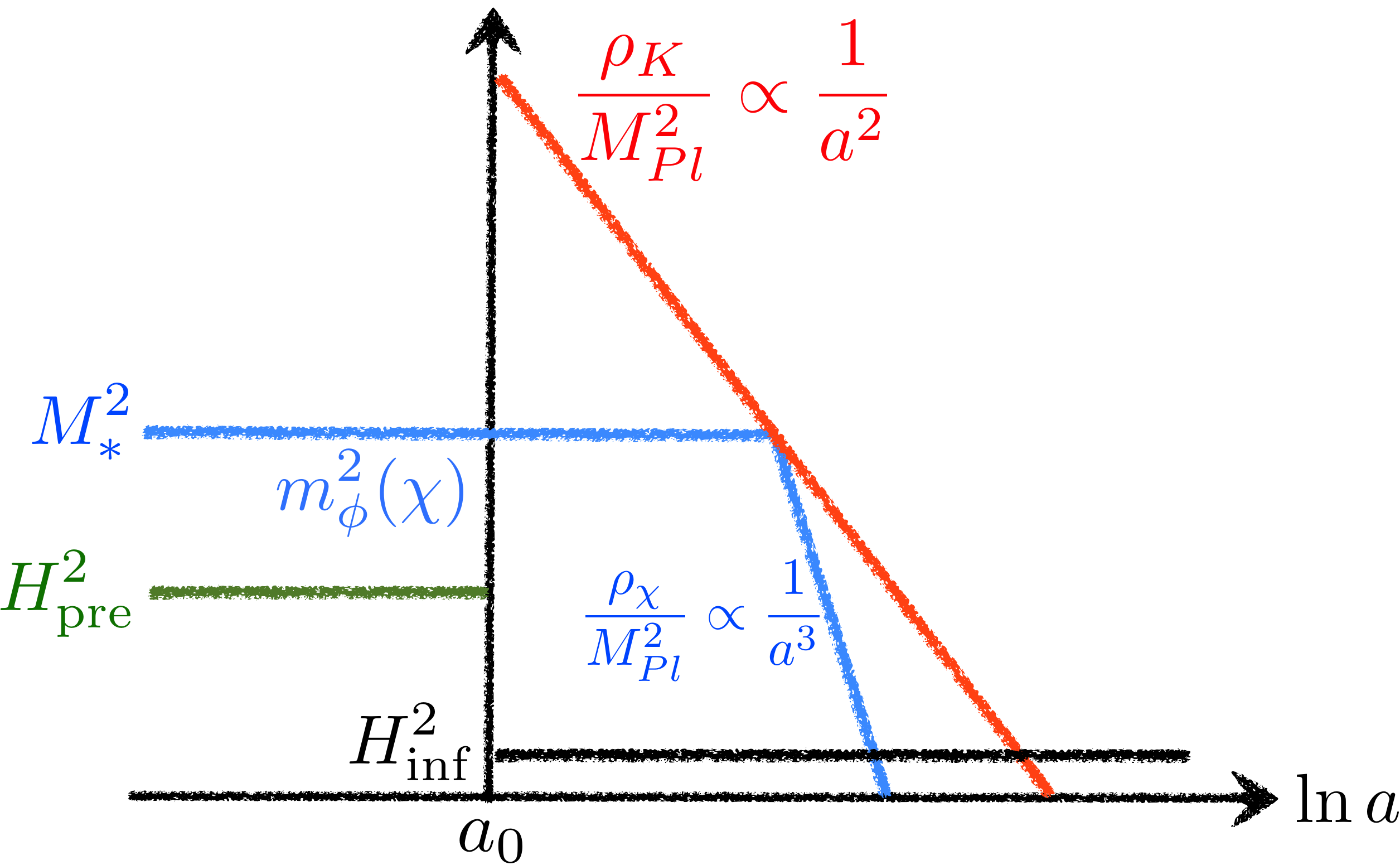}
\caption{The evolution of inflaton's Hubble induced mass.}
\label{fig:HubbleInducedMass}
\end{center}
\end{figure}

Let us follow the dynamics of the singlet scaler field $\chi$ and the inflaton $\phi$ before the inflation starts.
The mass of the inflaton in general depends on the energy density of the universe. Its evolution is summarised in Figure~\ref{fig:HubbleInducedMass}.

When the singlet field is at its metastable vacuum $\chi_{\rm pre}$, the potential energy $V_\chi$ of the singlet field dominates and the universe is in a precedent inflationary expansion. In the meanwhile, the inflaton acquires a Hubble induced mass and can be driven toward $\phi=0$. For example, in supergravity 
when the potential energy is dominated by the potential of the moduli field $V_\chi$, the potential includes
\beq
V \supset  \frac{V_\chi}{M_*^2} |\phi|^2,
\eeq
where $M_*$ is the cutoff scale.
We expect that $V_\chi \sim M_*^4$, and hence the Hubble induced mass of the inflaton, $m_\phi(\chi)$, during the precedent inflation is as large as $M_*$.
The similar is true for non-supersymmetric theories. We expect a coupling of the form
\beq
M_*^2 f(\frac{\chi}{M_*})\phi^2,
\eeq
where $f$ is some function, leading to the Hubble induced mass of $\calO(M_*)$. The Hubble scale $H_{\rm pre}$ is on the other hand of $\calO(M_*^2/ \Mp) \lesssim M_*$.
If the Hubble induced mass is positive, the inflaton is driven toward $\phi =0$.

After the quantum tunneling of the singlet field $\chi$ (denoted as $a_0$ in Figure \ref{fig:HubbleInducedMass}), the universe is dominated by the curvature energy density $\rho_K$ and the singlet field $\chi$ is fixed by the Hubble friction. Because the Hubble induced mass $m_\phi^2$ is proportional to $\rho_\chi$, we have $m^2_\phi \lesssim H^2 =  \rho_K/3\Mp^2$ right after the tunneling. Note that the curvature energy density alone does not give a Hubble induced mass term.%
\footnote{The coupling $|\phi|^2 R$ , where $R$ is the Rich scalar, gives a Hubble induced term through a potential energy of the universe.}
As the universe expands, $\rho_K$ decreases and when it becomes smaller than $m_\chi^2 \Mp^2 \sim M_*^2 \Mp^2$, the singlet field $\chi$ starts to roll down to the global minimum and oscillates.%
\footnote{When $\rho_K$ is larger than $M_*^2 \Mp^2$, the inverse of the size of the horizon exceeds the cut off scale $M_*$ and the validity of the effective field theory is questionable. The discussion here is applicable even if $\rho_K$ after the tunneling is as small as $M_*^2 \Mp^2$.}
At this point the mass of the inflaton is as large as the Hubble scale. However, since the energy density of the singlet field decreases as $a^{-3}$, the mass of the inflaton $m_\phi \sim \rho_\chi^{1/2} / \Mp$ does not exceed the Hubble scale of the expansion and hence the inflaton can be regarded as massless after the tunneling. When $\rho_K$ drops below the inflaton potential energy $H^2_{\rm inf} \sim \rho_\phi/\Mp^2$, the inflation begins.

Let us discuss the evolution of the fluctuation of the inflaton based on the above observation. 
Expanding the field into comoving momentum modes $\phi = \int \frac{d^3k}{(2\pi)^3} \phi_{\vec{k}} e^{i \vec{k} \cdot \vec{x}}$, the modes fluctuate with decreasing amplitude as the spacetime expands. During the precedent inflation era, after a mode exit the comoving horizon, $k=a H_{\rm pre}$, the amplitude is continually damped because of the Hubble induced mass and eventually vanishes in the superhorizon limit $aH_{\rm pre} \gg k$. Hence, right after tunneling, the superhorizon mode that has the largest amplitude is the one that exited the horizon right before tunneling. This mode has an amplitude
\beq
\label{eq:fluctuation}
\delta \phi_{\rm pre} \simeq H_{\rm pre} \sqrt{\frac{H_{\rm pre}}{m_\phi(\chi_{\rm pre})}} \simeq \frac{M_*^{5/2}}{\Mp^{3/2}}.
\eeq

The horizon after the tunneling resides inside the horizon before the tunneling. The comoving horizon $1/a H$ remains constant during curvature dominant era, and hence there is no horizon entrance nor exit. Thus inflaton fluctuation modes inside the horizon continue to be suppressed, while the long wavelength superhorizon modes are frozen as the inflaton is essentially massless during the curvature dominant era. Those frozen modes effectively work as the zero mode $\phi_i$ which obeys a Gaussian distribution with a zero mean and a variance $(\delta \phi_{\rm pre})^2$. $\phi_i$ is nothing but the initial condition of the new inflation.

In order to wipe out the curvature energy density and to have structures on the galaxy scale, the inflation needs to last long enough with an anthropic bound $\calN^{tot}_e \gtrsim \calN^{\rm ant}_e$. In \cite{Freivogel:2005vv}, it is found that in order to have typical galaxies being formed, the comoving Hubble scale at the time of photon decoupling should satisfy 
\beq
\frac{a_{dc} H_{dc}}{a_t H_t} >30,
\eeq
where the subscript $t$ denotes the time right after the quantum tunneling.\footnote{The effect of spatial curvature on structure formation is also discussed in \cite{Barrow:1982}.} With some manipulation we have
\beq
\frac{a_{dc} H_{dc}}{a_{ent} H_{ent}} \frac{a_* H_*}{a_{end} H_{end}} \frac{a_{end} H_{end}}{a_i H_i}
\frac{a_i H_i}{a_t H_t} > 30.
\eeq 
Here $a_{ent} H_{ent}$ denotes the coving Hubble scale of the horizon re-entrance of the CMB scale, which is equal to that of horizon exit $a_* H_*$. $a_{end} H_{end}$ and $a_i H_i$ are the comoving Hubble scales at the end and beginning of the inflation respectively. Because the period between the time right after quantum tunneling and  the beginning of inflation is curvature dominant, the comoving Hubble scale remains the same, i.e. $(a_i H_i)/(a_t H_t)$=1. We assume the Hubble scale during the inflation is nearly constant, $H_i \simeq H_* \simeq H_{end}$. Also, the comoving Hubble scale does not evolve much between the horizon re-entrance of the CMB scale and the photon decoupling, so $(a_{dc} H_{dc})/(a_{ent}H_{ent})\simeq 1$. Putting everything together, we then have a constraint
\beq
\calN^{tot}_{e} > \calN^{\rm ant}_e \simeq  \calN^*_e + 3.4\,,
\eeq
where $\calN^*_e$ is the number of e-foldings between the end of the inflation and the horizon exits of the CMB pivot scale.

For potentials of a new inflation type, the initial field value $\phi_i$ must be close to zero  to have long enough inflation. For a large enough $M_*$, $\delta \phi_{\rm pre}$ is larger than the required initial field value and hence some tuning of the initial field value is required.
As $\phi_i$ obeys a Gaussian distribution, which is flat for small $\phi_i$, the probability distribution of $\phi_i$ in the region of interest is approximately uniform;
\beq
\calP_{\phi_i} d \phi_i \propto d\phi_i.
\eeq

The anthropic constraint $\calN^{tot}_e \gtrsim \calN^{\rm ant}_e$ leads to the upper bound $\phi_i <\phi_{\rm ant}$, where $\phi_{\rm ant}$ is the field value of the inflaton such that the number of e-foldings after the inflaton pass the field value is $\calN^{\rm ant}_e$.

The fact that the initial condition $\phi_i$ has a probability distribution over a certain range instead of having to start at $\phi_i \simeq 0$ plays an important role to solve the $\eta$-problem in new inflation. Particularly, as now the inflaton tends to start from an initial condition away from zero, the anthropic constraints $\calN^{tot}_e > \calN^{\rm ant}_e$ requires the potential around the origin to be flatter. The $\eta$ parameter is biased toward smaller values after the anthropic constraint is imposed. On the contrary, for a small enough $M_*$ so that $\phi_i \simeq 0$ is forced, then the inflation can easily last longer than $\calN^{\rm ant}_e$ e-folds and the anthropic constraints on $\calN^{tot}_e$ plays no significant role. We will see this point quantitatively in the following. 

\subsection{A Supersymmetric New Inflation Model}
In the appendix we study a new inflation model with $Z_2$ symmetry, assuming that the parameters of the potential are uniformly distributed. We find that the resultant $P_\zeta$ is strongly biased toward a large value, and the observed one is probabilistically disfavored. Here we in stead investigate a supersymmetric model where it is sensible that the parameters of the model, including the scale of the inflation, obey distributions different from uniform ones. We expect that for certain distributions of the parameters, $P_\zeta$ is biased toward small ones.
In particular, we consider an $R$-symmetric single field new inflation model~\cite{Kumekawa:1994gx,Izawa:1996dv,Harigaya:2013pla} with a discrete $R$-symmetry $Z_{2N}$ is present and the superpotential
\beq
W= v^2 \Phi  - \frac{g}{N+1} \Phi^{N+1}, \label{SUSYW}
\eeq
where $\Phi$ is a chiral superfield while $v$ and $g$ are constants. Here and here after, we work in the unit where the reduced Planck scale is unity.
The \Kahler potential is
\beq
K = \Phi^\dagger \Phi + \frac{1}{4} k (\Phi^\dagger \Phi)^2 \cdots, \label{SUSYK}
\eeq
where the ellipses denote higher order terms that are irrelevant to the inflationary dynamics. From Eqs.(\ref{SUSYW}) and (\ref{SUSYK}), the potential of the scalar component of $\Phi$ which we call $\varphi$ is given by
\begin{align}
V(\varphi) & = |v^2 - g \varphi^N|^2 - k v^4 |\varphi|^2 + \cdots \nonumber \\
		& = v^4 - k v^4 |\varphi|^2 - \big(g v^2  \varphi^N + \text{ h.c.}  \big) + \cdots.
\end{align}
In terms of the radial and angular components, $\varphi = \frac{\phi}{\sqrt{2}}e^{i \theta}$, the potential can be rewritten as 
\beq
V = v^4 - \frac{1}{2} k v^4 \phi^2  - \frac{g}{2^{ \frac{N-2}{2}}} v^2 \phi^N \cos(N \theta).
\eeq

For simplicity we assume that the inflaton has an initial condition around $\theta = 0$ mod $2\pi/N$ (which are minima along the angular direction) and focus only on the radial direction.

In the appendix we study general new-inflation-type models with $Z_2$ symmetry.
Here the resulting potential has the form of Eq.(\ref{nmModel}) without the $c_m$ perturbation term. Using Eqs.(\ref{nmns}) and (\ref{nmPzeta}) with $a= v^4$, $b= k v^4$, $c_n = g v^2/2^{(N-2)/2}$ and $c_m=0$, we have
\begin{align}
n_s \simeq 1 - 2k -2N(N-1)k \fN^*  \quad, \label{nsSUSY} \\
P_\zeta = \frac{\Big[ g^2 v^{4(N-3)} k^{-2(N-1)} {\fN^*}^{-2} \Big]^{\frac{1}{N-2}}}{24 \pi^2 \left(1 + N \fN^* \right)^2} , \label{SUSYPzeta}
\end{align}
where
\beq
\fN \equiv \fN(k, \calN_e) = \frac{1}{N} \frac{1}{\left( \left[1+(N-1) k \right] e^{(N-2) k \calN_e} -1 \right)} 
\eeq
and $\fN^* = \fN(k, \calN_e^*)$ in which $\calN^*_e$ is the number of e-folds between the horizon-exit of the CMB scale and the end of inflation.

\begin{figure}[tb]
\begin{center}
\includegraphics[width=0.45\textwidth]{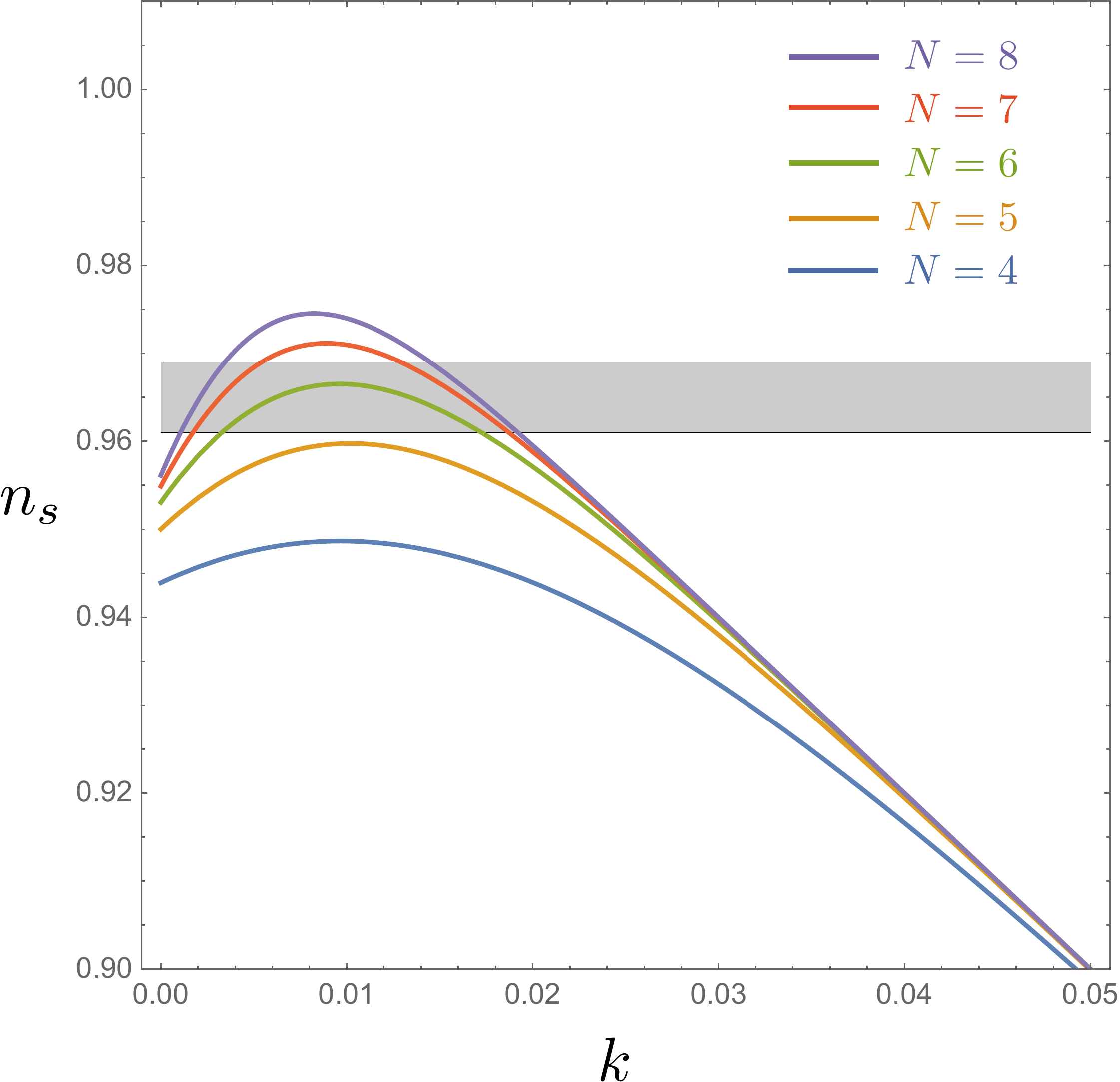}
\caption{The $n_s-k$ plot for $N=4, 5, 6, 7$ and 8, with $\calN^*_e =52$.}
\label{fig:SUSYns684}
\end{center}
\end{figure}

The spectral index $n_s$, as given in Eq.(\ref{nsSUSY}), is a function of the parameter $k$, and the number of e-folds $\calN^*_e$ between the horizon exit of the CMB scale and the end of inflation. Assuming instant reheating, this is determined by the inflation energy scale,
\beq
\calN^*_e \simeq 62 + \ln\left(\frac{v}{\Mp} \right).
\eeq
Using Eq.(\ref{SUSYPzeta}) with $k=0.01$ and the observed $P_\zeta$ to get an estimate on $v$, we have $\calN^*_{e} \simeq 52$. Setting $\calN^*_e = 52$, we plot the spectral index $n_s$ as a function of $k$ in Figure \ref{fig:SUSYns684}. One can see that $N=4$ and 5 are ruled out by the observation, while $N=6$  fits the observation very well in a certain region of $k$. Even if we relax the assumption of instant reheating, such that $\calN^*_e<52$, the maximum $n_s(k)$ of $N=6$ case lies in the observational allowed region unless the reheating temperature is very small. In below we frequently use $N=6$, i.e.~the model with discrete $Z_{12}$ $R$-symmetry, as a reference point.

\subsection{Probability Distribution of the Observables}
\label{sec:pdf}

A natural question now arises: what is the probability for $k$ to lie in the region that yields observationally allowed $n_s$? From Figure \ref{fig:SUSYns684} we see that the observed $n_s$ requires $k$ to be of the order of 0.01, while in general $k$ would be much larger. Note that the slow-roll parameter $\eta$ is related to $k$ by $\eta = -k$. Therefore, making the observed spectral index $n_s \sim 0.96$ probabilistically favorable is equivalent to solving the $\eta$-problem, and this requires a probability distribution that biases  toward small $k$.

To investigate the probability distribution function of the observables, we need to first make assumptions on the probability distribution of the Lagrangian parameters. As $k$ is a coupling in the \Kahler potential, it would be natural that $k$ obeys a uniform probability distribution $\calP(k) dk \propto dk$.
The parameters $v^2$ and $g$ are superpotential couplings and may obey distributions different from uniform ones. Furthermore, those parameters are related with the vacuum expectation value (vev) of the superpotential $W_0 \equiv \vev{W} \sim v^{2(1 + 1/N)} g^{-1/N}$, which is related with the electroweak scale and the cosmological constant.
Let us start from the distribution of $v^2$, $g$, the supersymmetry breaking scale $F$ and the $\mu$ term of the electroweak Higgs,
\beq
d v^2 v^{2p'}\times dg g^{q'} \times dF F^{r} \times d\mu \mu^{s}.
\eeq
If a parameter is given by a dimensional transmutation, the index ($p',q',r,s$) of the distribution is $-1$, while it is $1$ if the parameter is a complex parameter biased toward a large value.
The electroweak scale $v_{\rm EW}^2$ is given by
\beq
v_{\rm EW}^2 \simeq F^2/ M_{\rm med}^2 - \mu^2,
\eeq
where $M_{\rm med}$ is the mediation scale of the supersymmetry breaking. We assume that the electroweak scale must be in a certain range close to the observed one%
\footnote{This assumption is not crucial for our discussion. Without the anthropic constraint on the electroweak scale, the distribution is given by Eq.(\ref{eq:dist-pre}) with $s=1$.}
as is argued in~\cite{Agrawal:1997gf,Hall:2014dfa},
\beq
c v_{\rm EW,obs}^2 <v_{\rm EW}^2 < c' v_{\rm EW,obs}^2,
\eeq
where $c$ and $c'$ are constants which we do not have to specify. 
The scanning over the $\mu$ parameter yields
\beq
\int_{c v_{\rm EW,obs}^2 <v_{\rm EW}^2 < c' v_{\rm EW,obs}^2} d\mu \mu^{s} \simeq v_{\rm EW,obs}^2 \left( \frac{F}{M_{\rm med}}\right)^{s-1} \propto F^{s-1},
\eeq
where we have used $F/M_{\rm med} \simeq \mu \gg v_{\rm EW,obs}$ as is suggested by the non-discovery of supersymmetric particles so far. 
The cosmological constant is given by
\beq
\rho_\Lambda = F^2 - 3 W_0^2 \Mp^2,
\eeq
where $W_0$ is the vev of the superpotential. A change of variables gives  gives
\beq
\int d F F^{r + s-1} \propto W_0^{r + s - 2} d\rho_\Lambda.
\eeq
Here we have used $F^2 \simeq 3 W_0^2 \Mp^2 \gg \rho_\Lambda$. We omit the measure $d \rho_\Lambda$, which leads to the uniform distribution of the cosmological constant, in the following.
Using the relation $ W_0 \sim v^{2(1 + 1/N)} g^{-1/N}$, the distribution of $v^2$ and $g$ are given by
\beq
\label{eq:dist-pre}
dv^2 v^{2p}\times dg g^q,~~p = p' + (1 + \frac{1}{N}) (r + s-2),~~q = q' - \frac{1}{N} (r + s-2).
\eeq
For $-1 \leq p',q',r,s \leq 1$, a wide range of $(p,q)$ can be obtained. 

\begin{figure}[tb]
\begin{center}
\includegraphics[width=0.8\textwidth]{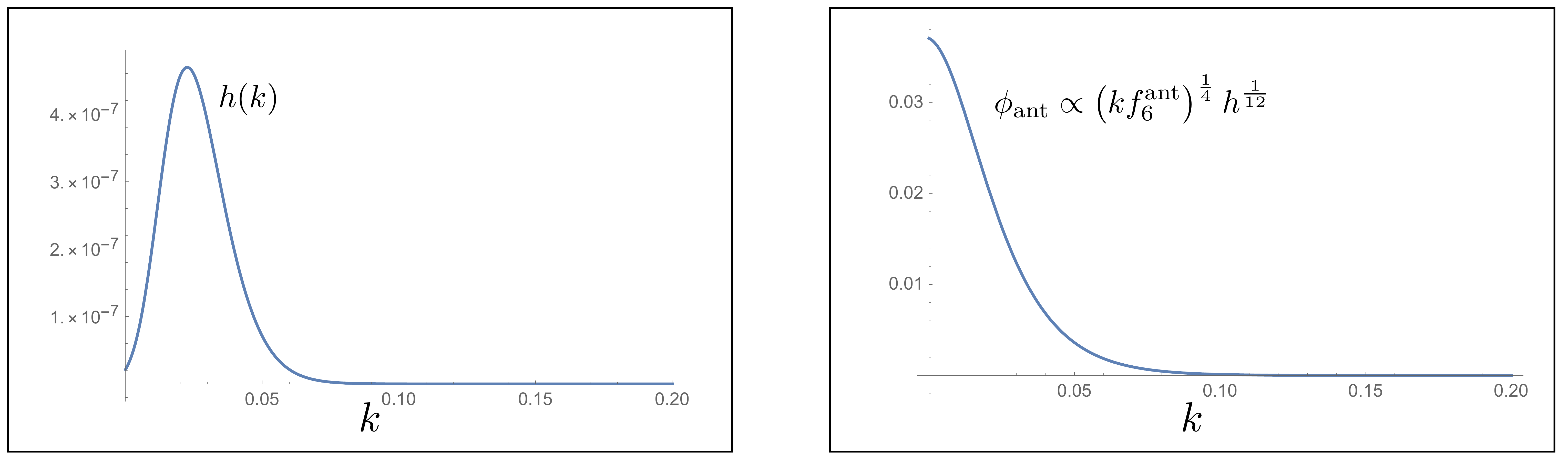}
\caption{The function $h(k)$ (left) and the contribution of $\phi_{\rm ant}$ to the probability distribution of $k$ (right). We take $N=6$ and $\calN^*_e = 52$.}
\label{fig:phi60contribution}
\end{center}
\end{figure}

Putting everything together, the probability distribution of the parameters to start with is
\beq
\int  d\phi_i\, dk\, dg\, \,dv^2 \,  \, g^q  \,v^{2p}.
\eeq
The distribution of the initial condition $\phi_i$ is uniform as discussed in Sec.\ref{subsec:InitialCondition}. If the anthropic bound $\phi_{\rm ant}$ is smaller than the amplitude of quantum fluctuation $\delta \phi_{\rm pre}$ given by Eq.(\ref{eq:fluctuation}), then the integration of $d\phi_i$ ranges from $0$ to $\phi_{\rm ant}$. On the other hand, if $\phi_{\rm ant} > \delta \phi_{\rm pre}$, the integration of $d\phi_i$ is capped by $\delta \phi_{\rm pre}$ and the anthropic constraint is no longer effective. In other words, integrating out $d\phi_i$ yields
\begin{align}
\int  \, dk\, dg\, \,dv^2 \,  \, g^q  \,v^{2p}  \phi_{\rm bound}, ~~~
\phi_{\rm bound} \equiv \min[\phi_{\rm ant}, \delta \phi_{\rm pre}].
\end{align}
Using Eqs.(\ref{phiNe}), (\ref{nmcn0}) and (\ref{SUSYPzeta}), the field value $\phi_{\rm ant}$  is given by
\begin{align}
\phi_{\rm ant} & = \sqrt{2}\, g^{-\frac{1}{N-3}} k^\frac{2}{N-3} {\fN^*}^\frac{1}{(N-3)(N-2)} \left[ 24 \pi^2 \left(1+ N \fN^* \right)^2 \right]^\frac{1}{2(N-3)} {\fN^{\rm ant}}^\frac{1}{N-2} P_\zeta^\frac{1}{2(N-3)} \nonumber \\
& = \sqrt{2} g^{-\frac{1}{N-3}} {\fN^{\rm ant}}^\frac{1}{N-2} k^{\frac{1}{N-2}} h^{\frac{1}{(N-3)(N-2)}} P_\zeta^{\frac{1}{2(N-3)}}, \label{phiant}
\end{align}
where we have defined
\beq
h(k) \equiv k^{N-1} \fN^* \left[24 \pi^2 (1+N \fN^*)^2 \right]^{\frac{N-2}{2}}
\eeq
for future convenience. It is instructive to understand the behavior of $h(k)$ and the contribution of $\phi_{\rm ant}$, if $\phi_{\rm ant} < \delta \phi_{\rm pre}$, to the final probability distribution of $k$ which we plot in Figure~\ref{fig:phi60contribution}. Most importantly, we can see that $\phi_{\rm ant}$ gives a bias toward small $k$, which is the key of solving the $\eta$-problem and making observed spectral index $n_s$ probabilistically favorable.

Using the relation between $v^2$ and $P_\zeta$ derived from Eq.(\ref{SUSYPzeta}),
\beq
v^2 = g^{-\frac{1}{N-3}} h^\frac{1}{N-3} P_\zeta^\frac{N-2}{2(N-3)}, \label{v2SUSY}
\eeq
we can perform a change of variable to obtain the probability distribution in terms of cosmic perturbation $P_\zeta$. For the parameter region where $\phi_{\rm ant} < \delta \phi_{\rm pre}$, we have
\begin{align}
&\int  \, dk\, dg\, \,dv^2 \,  \, g^q  \,v^{2p}  \phi_{\rm ant}\\
 = &\frac{(N-2)}{\sqrt{2} (N-3)} \int dk\, \frac{dP_\zeta}{P_\zeta} \, dg \, g^{\frac{q(N-3)-p-2}{N-3}} 
h^{\frac{(p+1)}{(N-3)}} P_\zeta^{\frac{(N-2)(p+1)}{2(N-3)}} \left[\left(k \fN^{\rm ant} \right)^\frac{1}{N-2} h^{\frac{1}{(N-3)(N-2)}} P_\zeta^\frac{1}{2(N-3)} \right], \label{SUSYgnonIntegrate}
\end{align}
where we have left the contribution from $\phi_{\rm ant}$ to $k$- and $P_\zeta$-distribution in the square brackets for future convenience. We now need to integrate out $g$ to obtain the final probability distribution. When $q(N-3)-p+N-5 \neq 0$, the integration yields
\beq
 \frac{(N-2)}{\sqrt{2} \left[ q(N-3)-p+N-5 \right]}  \int dk\,
\frac{dP_\zeta}{P_\zeta} \left[ g^{\frac{q(N-3)-p+N-5}{(N-3)}}   \right]^{g_{max}}_{g_{min}}
h^{\frac{(p+1)}{(N-3)}} P_\zeta^{\frac{(N-2)(p+1)}{2(N-3)}} \Big[\cdots \Big] \label{SUSYgIntegrated}
\eeq
where the ellipses in the square brackets represent the contribution from $\phi_{\rm ant}$ as in Eq.(\ref{SUSYgnonIntegrate}). The lower cutoff of the integral $g_{min}$ is given by Eq.(\ref{v2SUSY}) with the natural requirement $v^2 \leq M_*^2$, i.e.~the energy scale $v$ should not be larger than the cut off scale. Particularly, we have
\beq
g_{min} = h P_\zeta^{\frac{N-2}{2}}M_*^{-(N-3)} \ll 1.
\eeq
On the other hand,  $g_{max}$ is determined by the cutoff scale $M_*$. After restoring $\Mp$ and $M_*$ to the superpotential, we have
\beq
W \supset - \frac{1}{N+1} \frac{g}{\Mp^{N-2}} \Phi^{N+1} \equiv - \frac{1}{N+1} \frac{c_g}{M_*^{N-2}}\Phi^{N+1}.
\eeq
Assuming the dimensionless coupling $c_g$  is bounded by unity, the coupling $g$ is bounded by
\beq
g < g_{max} = \left(\frac{\Mp}{M_*} \right)^{N-2}.
\eeq
 Therefore, if $q(N-3)-p+N-5 > 0$, the $g-$integration merely gives a proportional constant and does not affect the probability distribution of $k$ and $P_\zeta$. We therefore have
\beq
\calP_{\rm inf}(k, P_\zeta) dk\,dP_\zeta =  \calP_k \calP_{P_\zeta} dk\,dP_\zeta   = \calP_k P_\zeta \calP_{P_\zeta} dk\,d{\rm ln}P_\zeta
\eeq
where $\calP_{\rm inf}$ is the probability distribution from inflationary dynamics with
\begin{align}
\calP_k & \propto h^{\frac{(p+1)}{(N-3)}} \left[ \left(k \fN^{60} \right)^\frac{1}{N-2} h^{\frac{1}{(N-3)(N-2)}} \right], \label{SUSYPkpcase}\\
P_\zeta\calP_{P_\zeta} & \propto P_\zeta^{\frac{(N-2)(p+1)}{2(N-3)}} P_\zeta^\frac{1}{2(N-3)}.\label{SUSYPPzetapcase}
\end{align}
The quantity $P_\zeta\calP_{P_\zeta}$ can be understood as the relative probability to obtain the curvature perturbation of $\calO({\cal P}_\zeta)$.
When  $q(N-3)-p+N-5 = 0$, Eq.(\ref{SUSYgIntegrated}) does not apply and the integration over $g$ yields a logarithmic contribution $\ln(g_{min})$ instead. This logarithmic contribution changes the probability distribution of $k$ and $P_\zeta$ only slightly, and we may use Eqs.(\ref{SUSYPkpcase}) and (\ref{SUSYPPzetapcase}) as a good approximation. 

If $\phi_{\rm ant} > \delta \phi_{\rm pre}$, then $\phi_{\rm ant}$ plays no role and the integration of $\phi_i$ and $g$ merely yields a proportional constant  which do not affect the probability distribution of the observables:
\beq
\int  \, dk\, dg\, \,dv^2 \,  \, g^q  \,v^{2p}  \phi_{\rm bound}
\propto 
\int dk\, \frac{dP_\zeta}{P_\zeta} 
h^{\frac{(p+1)}{(N-3)}} P_\zeta^{\frac{(N-2)(p+1)}{2(N-3)}}.
\eeq
Thus for $\phi_{\rm ant} > \delta \phi_{\rm pre}$, we obtain
\begin{align}
\calP_k & \propto h^{\frac{(p+1)}{(N-3)}}, \label{SUSYPkpcaseNOant}\\
P_\zeta\calP_{P_\zeta} & \propto P_\zeta^{\frac{(N-2)(p+1)}{2(N-3)}}.
\label{SUSYPPzetapcaseNOant}
\end{align}
Note that in the parameter region $q(N-3)-p+N-5 \geq 0$, the probability distribution is parametrized only by $p$ but not by $q$.

In the parameter region where $q(N-3)-p+N-5 < 0$, the integration over $g$ is dominated by the lower cutoff contribution, where $g = g_{min}\ll 1$. Recall that $\phi_{ant} \propto g^{-1/(N-3)}$, and hence for such a small $g$, $\phi_{ant}$ is much larger than $\Mp$. This means the inflaton field value at the CMB scale will also be much larger than $\Mp$, and therefore the assumption of small field inflation breaks down.

\begin{figure}[tb]
\begin{center}
\includegraphics[width=\textwidth]{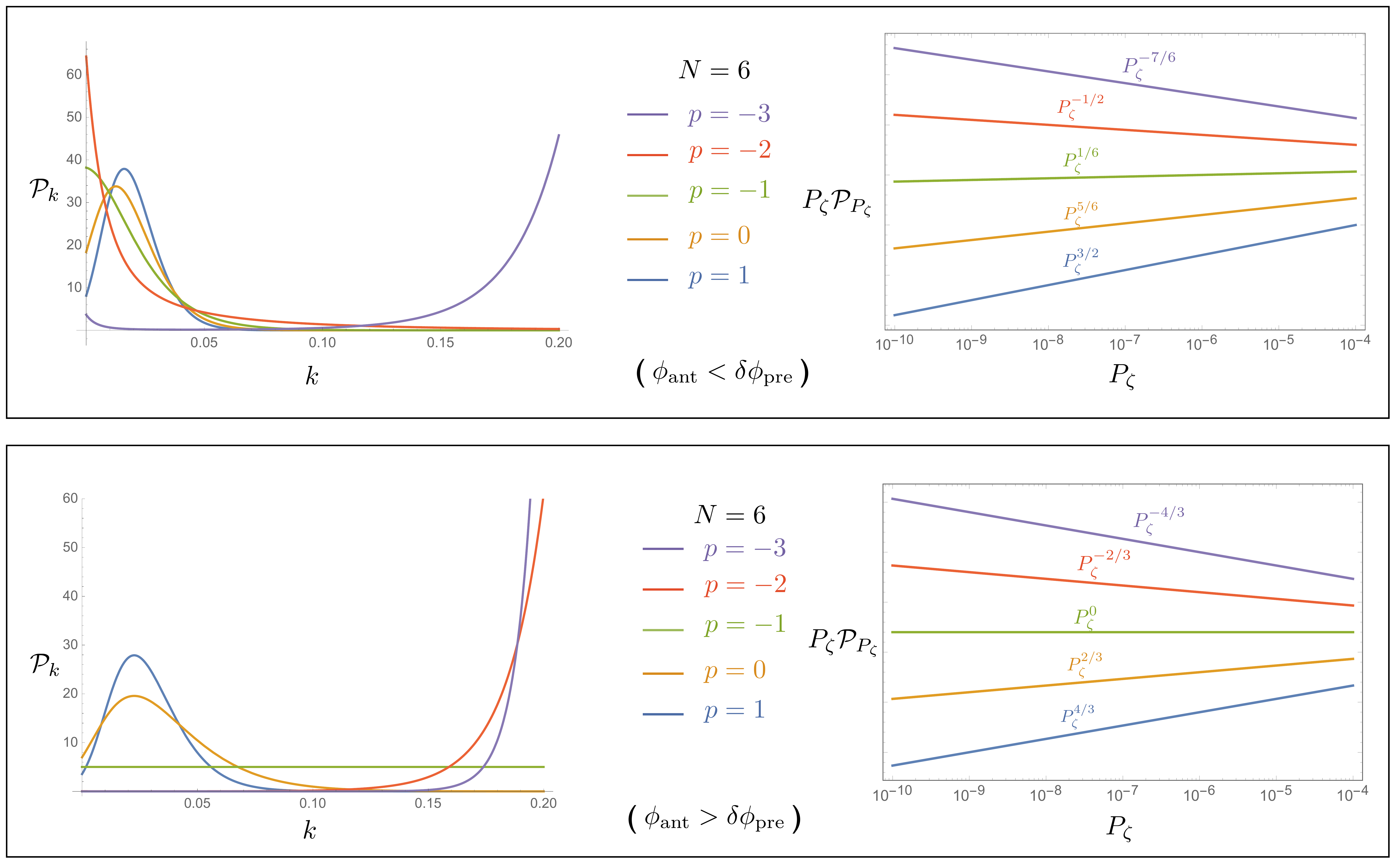}
\caption{ The probability distribution functions in the parameter region $q(N-3)-p+N-5 \geq 0$ when $\phi_{\rm ant} < \delta \phi_{\rm pre}$ (upper panel) and $\phi_{\rm ant} > \delta \phi_{\rm pre}$ (lower panel), respectively.  We take $N=6$ and $\calN^*_e = 52$.}
\label{fig:PDFp}
\end{center}
\end{figure}

In Figure \ref{fig:PDFp} we plot the distribution function using Eqs.(\ref{SUSYPkpcase}), (\ref{SUSYPPzetapcase}), (\ref{SUSYPkpcaseNOant}) and (\ref{SUSYPPzetapcaseNOant}) in the parameter region $q(N-3)-p+N-5 \geq 0$ for both  $\phi_{\rm ant} < \delta \phi_{\rm pre}$ and  $\phi_{\rm ant} > \delta \phi_{\rm pre}$. Let us first look at the upper left panel of Figure \ref{fig:PDFp}. We see that $\calP_k$ is largely suppressed for large $k$ as long as  $p \geq -2$. To understand how the parameter $p$ alters the probability distribution at large $k$, note that the functions $\fN$ and $h$ behave as
\begin{align}
\fN & \propto k^{-1} e^{-(N-2) k\calN_e} \label{fNlargek}\\
h & \propto k^{N-2} e^{-(N-2) k \calN_e} \label{hlargek}
\end{align}
when $k\to \infty$, and hence
\beq
\calP_k \propto k^{\frac{(N-2)(p+1)+1}{N-3}} e^{- \frac{(N-2)(p+2)}{N-3} k\calN_e}
\eeq
for large $k$. It is therefore clear that in order to solve the $\eta$-problem, one requires $p\geq -2$ so that the distribution is suppressed for large $k$.

The behavior of $\calP_k$ for negative $p$ can be understood as follows. Recall that the primordial perturbation is given as
\beq
P_\zeta \sim \frac{V}{\epsilon} \sim \frac{v^4}{k^2 \phi^2_{\rm cmb}}.
\eeq
where $\phi_{\rm cmb}$ is the inflaton field value when the CMB scale exited the horizon. Therefore, for a fixed $P_\zeta$, smaller $v^2$ requires smaller $ k \phi_{\rm cmb}$. For a given e-folds $\calN^*_e$, the field value $\phi_{\rm cmb}$ is in fact related to the parameter $k$. For larger $k$, the potential is steeper and hence $\phi_{\rm cmb}$ has to be smaller to maintain the same number of e-folds. (This relation is explicitly shown in Figure \ref{fig:phik46} in the appendix.)
The full $k$-dependence of the denominator $k^2 \phi^2_{\rm cmb}(k)$ is thus nontrivial, and is worked out in Eq.(\ref{phiNe}). It turns out that when $k$ decreases, $k^2 \phi^2_{\rm cmb}(k)$ increases. Therefore, for a given $P_\zeta$, if $v^2$ is biased toward small values as when $p$ is negative, then $k$ is biased toward large values. This is why large $k$ is favored when $p$ is too negative, where the bias toward small $k$ from $\phi_{\rm ant}$ is defeated.

\begin{figure}[tb]
\begin{center}
\includegraphics[width=0.9\textwidth]{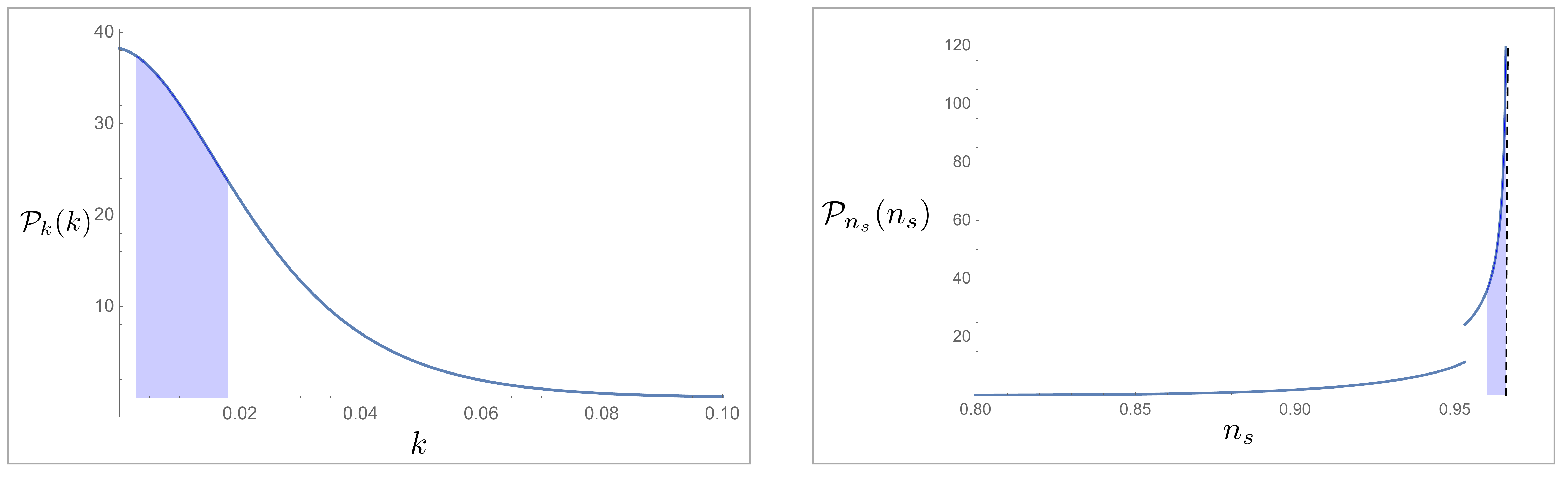}
\caption{The normalized probability distribution function for $k$ (left) and $n_s$ (right). The shaded areas corresponds to the parameter region where $n_s>0.96$. }
\label{fig:SUSYknsdis}
\end{center}
\end{figure}

So far we have only discussed the probability distribution in terms of $k$ but not the observable $n_s$. Since the spectral index $n_s$ is a function of $k$ only, as given in Eq.(\ref{nsSUSY}), the probability distribution of $k$ is sufficient to give the probabilistic information of $n_s$. That being said, the probability distribution of $n_s$ displays an important feature of $R$-symmetry new inflation which we now discuss. To this end, we perform a change of variable from $k$ to $n_s$,
\beq
\int dk \,\calP_k = \int dn_s \left|\frac{\p k}{\p n_s} \right| \calP_k(k(n_s)) \equiv \int dn_s \,\calP_{n_s}.
\eeq
As shown in Figure \ref{fig:SUSYns684}, the function $n_s(k)$ is not monotonic and has a maximum $n^{max}_s$ very close to the observed value $n_s\simeq 0.965$. These two properties have important implications in $\calP_{n_s}$ we show in Figure \ref{fig:SUSYknsdis}. The first feature of $\calP_{n_s}$ is the jump due to non-monotonicity of $n_s(k)$. The second, and probably more important, feature of $\calP_{n_s}$ is that $n_s$ will never reach $n_s=1$ and $\calP_{n_s}$ diverges at $n^{max}_s$ because the Jacobian factor $\left|\frac{\p k}{\p n_s} \right|$ diverges at $n^{max}_s$. Once the probability distribution of large $k$ ($n_s\ll 1$) is suppressed due to the probabilistic nature of initial field value $\phi_i$, in $R$-symmetry new inflation we not only can explain why $n_s$ is very close to one, but can also predict an $n_s\neq 1$ that is near the observed value $n_s\simeq 0.965$.
For $\calN^*_e=52$ and $p=-1$, the probability for $0.96<n_s < n^{max}_s $ is 
\beq
P_{0.96<n_s < n^{max}_s } = \frac{\displaystyle \int^{n^{max}_s}_{0.96} \,\calP_{n_s} \,dn_s}{\displaystyle\int^{n^{max}_s}_{-\infty} \,\calP_{n_s} \,dn_s} \simeq 0.48 \,,
\eeq
and the probability distribution diverges at $n_s^{max} \simeq 0.966$. Note that $p=-2$  yields a similar result.

Move on to the probability distribution of $P_\zeta$, the upper right panel of Figure \ref{fig:PDFp} shows that $P_\zeta$ is biased toward smaller values for a sufficiently negative $p$. This is simply because $P_{\zeta}$ is proportional to the inflation scale, and hence a bias toward small $v^2$ results in a bias toward small $P_\zeta$. For $p = -2$, the perturbation $P_\zeta$ is biased toward small value strongly. For $p=-1$,
$P_\zeta$ is biased toward large values only mildly.
The power of $P_\zeta \calP_{P_\zeta}$, as shown in Eq.(\ref{SUSYPPzetapcase}), is given by
\beq
\frac{p(N-2)+(N-1)}{2(N-3)} = \frac{(p+2)(N-2)}{2(N-3)}-\frac{1}{2}.
\eeq
In order to solve the $\eta$-problem simultaneously, we need $p\geq -2$. The most negative power we can get for the probability of $P_\zeta$ from inflationary dynamics is then $-1/2$.

As we mentioned in the introduction, the anthropic consideration on the post-inflation dynamics gives an additional bias on $P_\zeta \calP_{P_\zeta}$ that scales as $P_\zeta^{3/2}$ for small $P_\zeta$, and scales as $P_\zeta^{-1/2}$ for large $P_\zeta$ where the turning point is at $P_\zeta \sim 10^{-8}$. See Figure \ref{fig:PPzetaPost}.  Combining this with the contribution from inflationary evolution, we see that the power of $P_\zeta \calP_{P_\zeta}$ should be smaller or equal to $\frac{1}{2}$,
since otherwise $P_\zeta\sim O(1)$ is much more favored than $P_\zeta\sim 10^{-9}$. This requires
\beq
p\leq \frac{-2}{N-2}.
\eeq
Recall that in order to solve the $\eta$-problem, one requires $p\geq -2$. Hence, to  \textit{simultaneously} solve the $\eta$-problem and explain the smallness of perturbation power spectrum, we need
\beq
-2 \leq p \leq \frac{-2}{N-2}.
\eeq
The probability to obtain the observed value $P_\zeta \simeq 2.1 \times 10^{-9}$  is maximized when $p=-2$: With an anthropic bound on the density contrast $\delta \rho / \rho < \calO(10^{-4})$ from the property of galaxies, the probability is $O(10)$\%, which is reasonable.

To show the impact of the bias from $\phi_{\rm ant}$, in the lower panel of Figure \ref{fig:PDFp}, we show the distribution functions \textit{without} the $\phi_{\rm ant}$ contribution originated from the probabilistic nature of the initial field value $\phi_i$, which is the case when $\phi_{\rm ant} > \delta \phi_{\rm pre}$. Comparing with the upper panel, we see that this does not affect the distribution of $P_\zeta$ much. However, for the distribution of $k$, the probability for large $k$ is suppressed only for $p = 0$ and 1. For $p=-1$, without the additional suppression at large $k$ from $\phi_{\rm ant}$ as illustrated in the right figure of Figure \ref{fig:phi60contribution}, we have a uniform distribution in $k$ and hence it is more likely to find $k$ to be of order 1, instead of order 0.01.  Comparing both distributions in the lower panel of Figure \ref{fig:PDFp}, it is clear that without scanning the initial condition $\phi_i$, it is impossible to simultaneously solve the $\eta$-problem and explain the smallness of $P_\zeta$.

\begin{figure}[tb]
\begin{center}
\includegraphics[width=0.45\textwidth]{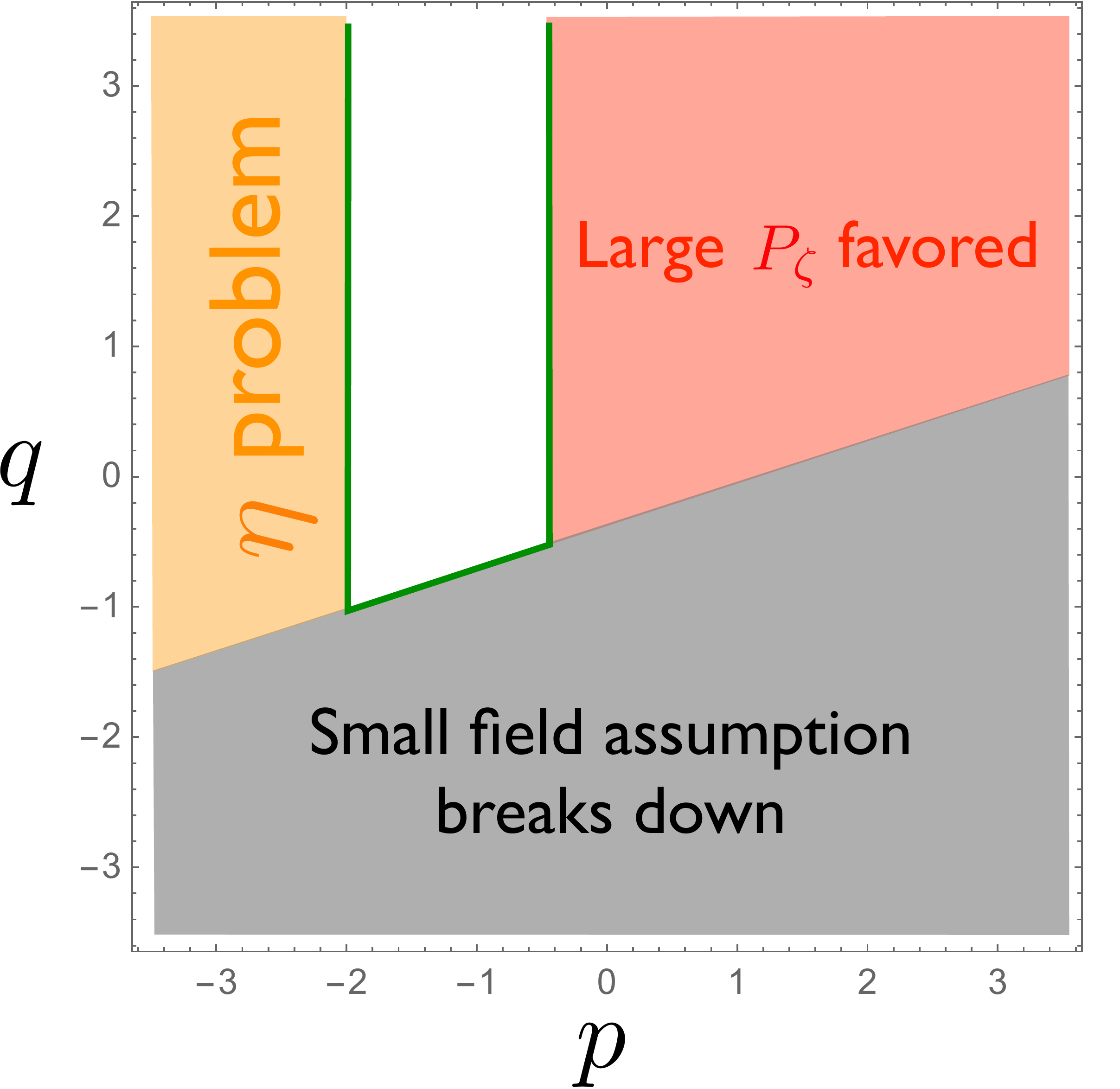}
\caption{Illustration of the result in $(p,q)$ parameter space for $N=6$. The white region represents the parameter space where both the $\eta$-problem and the smallness of cosmic perturbation can be explained.}
\label{fig:pqSpace}
\end{center}
\end{figure}

We summarize the discussion of the $(p,q)$ parameter space for $N=6$ in Figure \ref{fig:pqSpace}. The gray-shaded region is where $q(N-3)-p+N-5 < 0$ and the small field assumption breaks down; the red region is where the probability distribution of $P_\zeta$ biases toward large value even though the parameter $k$ tends to be small; the orange region is the opposite, where the $\eta$-problem persists despite the smallness of the perturbation power spectrum can be explained. In between the two regions we have parameter sets that can solve both problems. 
Those values of $(p,q)$ can be obtained by appropriate choice of $(p',q',r,s)$.

In claiming the existence of viable $(p,q)$, we assume the contribution to the distribution of $P_\zeta$ from the post-inflationary dynamics shown in Figure~\ref{fig:PPzetaPost}. If the power of the distribution at large $P_\zeta$ increases/decreases because of possible biases we have not considered, the white region in Figure~\ref{fig:pqSpace} shrinks/expands. The white region exists as long as the power is smaller than $1/2$.

\begin{figure}[tb]
\begin{center}
\includegraphics[width=0.9\textwidth]{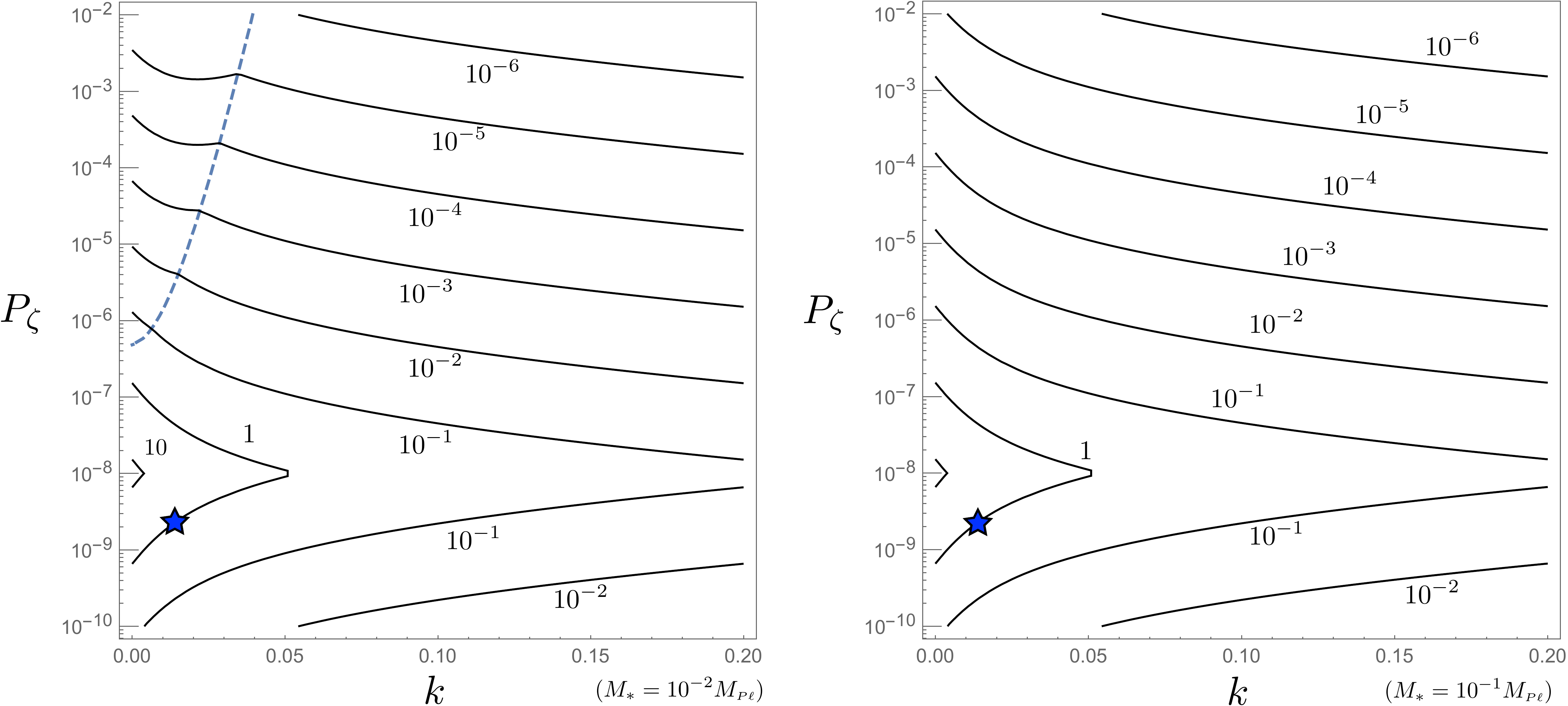}
\caption{The contour plot of $\frac{P_\zeta \calP_{\rm net}(k, P_\zeta) }{P_\zeta\calP_{\rm net}\big|_{k_{obs}, P_\zeta^{obs}}}$ for $N=6$, $p=-2$ and $M_* = 10^{-2}\Mp$ ($M_* = 10^{-1}\Mp$) on the left (right). To the right of the blue dashed line, $\phi_{\rm ant} < \delta \phi_{\rm pre}$ and hence the anthropic constraints $\phi_i <\phi_{\rm ant}$ contributes to the probability distribution. The blue star marks the point $(k_{obs}, P_\zeta^{obs})$}
\label{fig:PkSpace}
\end{center}
\end{figure}

It is worth emphasizing again that the viable parameter sets, the white region in Figure \ref{fig:pqSpace}, exists because of the probabilistic nature of the inflaton initial field value $\phi_i$. Without this contribution, the window between the red and orange regions is closed. 
We examine in which part of the parameter space is $\phi_{\rm ant} < \delta \phi_{\rm pre}$ and does the contribution of $\phi_{\rm ant}$ kick in. Recall that the amplitude of the quantum fluctuation $\delta \phi_{\rm pre}$ is proportional to $M_*^{5/2}$ as given in Eq.(\ref{eq:fluctuation}). On the other hand, $\phi_{\rm ant}$ also depends on $M_*$ through the superpotential coupling $g$. The probability distribution then has the form
\begin{align}
&\int  \, dk\, dg\, \,dv^2 \,  \, g^q  \,v^{2p} \, \phi_{\rm bound}(k, P_\zeta, g; M_*) \nonumber \\
 = &\frac{(N-2)}{\sqrt{2} (N-3)} \int dk\, \frac{dP_\zeta}{P_\zeta} \, dg \, g^{\frac{q(N-3)-p-1}{N-3}} 
h^{\frac{(p+1)}{(N-3)}} P_\zeta^{\frac{(N-2)(p+1)}{2(N-3)}} \phi_{\rm bound}(k, P_\zeta, g; M_*).
\end{align} 
The parameter region of interest is $q(N-3) - p + N-5 >0$. Therefore after integrating out $g$, the integration is dominated by $g_{max} = (\Mp/M_*)^{N-2}$, at which
\beq
\phi_{\rm bound}(k, P_\zeta; M_*) = \min \big[\phi_{\rm ant}(k, P_\zeta, g_{max}), \delta \phi_{\rm pre}\big].
\eeq
Combining with the contribution from the anthropic constraint discussed in the introduction, 
\beq
\calP_{\rm post}\left({P_\zeta}\right) = \min\left[(P_\zeta/10^{-8})^{3/2}, (P_\zeta/10^{-8})^{-1/2} \right],
\eeq
the net probability distribution $P_\zeta \calP_{\rm net}(k, P_\zeta)$ in the $(k, P_\zeta)$ space that includes both inflationary and post-inflationary dynamics is proportional to
\beq
h^{\frac{(p+1)}{(N-3)}} P_\zeta^{\frac{(N-2)(p+1)}{2(N-3)}}  \min \big[\phi_{\rm ant}(k, P_\zeta, g_{max}), \delta \phi_{\rm pre}\big]   \min\left[(P_\zeta/10^{-8})^{3/2}, (P_\zeta/10^{-8})^{-1/2} \right].
\eeq
The distribution $P_\zeta\calP_{\rm net}$ normalized with respect to $P_\zeta\calP_{\rm net}\big|_{k_{obs}, P_\zeta^{obs}}$ for $N=6$, $p=-2$ and $M_* = 10^{-2}\Mp$  is given in the left panel of Figure \ref{fig:PkSpace}, where $k_{obs}\simeq 0.0134$ and $P_\zeta^{obs}\simeq 2.2\times 10^{-9}$ are the observed value for the parameter $k$ and the cosmic perturbation $P_\zeta$ respectively and $(k_{obs},  P_\zeta^{obs})$ is marked by the blue star. To the right of the blue dashed line, $\phi_{\rm ant} < \delta \phi_{\rm pre}$ and the anthropic constraints $\phi_{\rm ant}$ plays an important role to solve the $\eta$-problem. The observed point $(k_{obs}, P_\zeta^{obs})$ lies deep inside the region and hence the proposed scenario can indeed explain the nearly scale invariant small cosmic perturbation. Note that the cusps at $P_\zeta \sim 10^{-8}$ originate from the turning point of $\calP_{\rm post}(P_\zeta) $, while the cusps at the blue dashed line are due to $\min \big[\phi_{\rm ant}(k, P_\zeta, g_{max}), \delta \phi_{\rm pre}\big] $. The distribution for $M_* = 10^{-1}\Mp$ is given in the right panel, where the blue dashed line is absent because $\delta \phi_{\rm pre} \propto M_*^{5/2}$ is always larger than $\phi_{\rm ant}$.

\section{Summary and Discussion}
\label{Sec:Discussion}
In this work we have investigated the typicality of the small and nearly scale-invariant perturbation
in the landscape. Anthropic consideration of the cosmological constant yields a probability of $P_\zeta$ that biases toward large perturbation, until the anthropic constraint due to the density of the galaxy kicks in. In order for the observed small $P_\zeta \sim 10^{-9}$ to be typical, the inflationary evolution has to give a bias toward small $P_\zeta$. Closeness of the spectral index to the unity should be also explained.

We consider the following scenario that naturally fits into the landscape scenario: The inflaton is coupled to a singlet scalar field that was initially trapped in a metastable vacuum and drove a precedent inflation. After the quantum tunneling of the singlet field, the universe became an open FRW universe dominated by the curvature energy density while the singlet field rolled down to a stable vacuum with negligible energy density. After a sufficient period of cosmic expansion, when the curvature energy density dropped below the potential energy of the inflaton, the inflation which explains the flatness of the universe and the cosmic perturbation occurred.

In this scenario, the inflaton field value is homogeneous inside the horizon because of the trapping during the precedent inflation. However, after the quantum tunneling the universe is curvature dominated and the trapping is no longer effective. As a result quantum fluctuation of long wavelength modes produced during the precedent inflation survive, which leads to a probabilistic nature of  the initial field value of the inflaton. As the inflaton tends to start from the an initial condition away from the origin, the anthropic lower bound on the total number of e-folding during inflation favors the inflaton potential flatter around the origin, namely a smaller $\eta$ parameter.

We investigated a supersymmetric new inflation model in detail. We find that for certain distributions of the parameters, the probability to obtain $P_\zeta \sim 10^{-9}$ is $\calO(10)$\%, while the observed $n_s$ is favored. We emphasize that both the model-building and the anthropic selection from the landscape play important roles in explaining the observed properties of the cosmic perturbation, $P_\zeta$ and $n_s-1$.
From the model-building side, the distribution function of the model parameters, which is not uniform owing to the supersymmetry and the $R$ symmetry, yields the distribution of $P_\zeta$ not biased toward large values. From the landscape side, the requirement of large enough number of e-foldings and the probabilistic nature of the initial inflaton field value set by the precedent inflation dynamics favor small $\eta$ parameter, thereby explaining the observed $n_s$.

The result is encouraging for the project on understanding the universe by the anthropic principle in the landscape. Further study is required toward this goal.
For instance, in this paper we assume the contribution to the distribution of $P_\zeta$ from the post-inflationary dynamics shown in Figure~\ref{fig:PPzetaPost}. As we comment in Sec.~\ref{sec:pdf}, our result holds qualitatively as long as the power of the distribution at large $P_\zeta$ is smaller than $1/2$. It will be important to investigate the distribution at large $P_\zeta$ more carefully, taking into account the effect of e.g.~the behavior of proto-galaxies.

\section*{Acknowledgement}
We thank Yasunori Nomura for useful discussions throughout the collaboration as well as comments on the manuscript. We also thank Hitoshi Murayama for helpful discussions that improve the presentation of the draft, as well as David Dunsky and Vijay Narayan for stimulating discussion about the manuscript.
This work was supported in part by the Director, Office of Science, Office of High Energy and Nuclear Physics, of the US Department of Energy under Contract DE-AC02-05CH11231(CIC, KH) and DE-SC0009988 (KH) as well as by the National Science Foundation under grants PHY-1316783 (KH) and PHY-1521446 (KH).

\appendix

\section{Fine-tuning in General New Inflation}\label{Sec:FineTuneGeneral}
In the appendix we study fine-tuning problems in general new-inflation-type models. The only symmetry we impose here is $Z_2$ symmetry, where the most generic potential is of the form
\beq
V(\phi) = \Mp^4 \left[V_0 + \sum_{n=1}^{\infty} c_{\scpt 2n} \left(\frac{\phi}{\Mp} \right)^{2n} \right]. \label{generalZ2}
\eeq
Hereafter we will work in the Planck unit where the reduced Planck mass $\Mp$ is set to unity. We study how much fine-tuning is required to yield the observed perturbation amplitude and spectrum. We assume that the probability distributions for the dimensionless coefficients $V_0$ and $|c_{\scpt 2n}|$'s are uniform between zero and one, and vanish outside this interval. Certainly, not all possible values of $c_{\scpt 2n}$ allow inflation, as slow-roll conditions 
\begin{align}
\epsilon & = \frac{1}{2} \left(\frac{V' }{V} \right)^2
\simeq \frac{1}{2} \left(\frac{\sum_{n=1}^\infty 2n\, c_{\scpt 2n} \phi^{2n-1}}{V_0} \right)^2, \label{epsiloncondition} \\[10pt]
\eta & = \frac{V''}{ V} \simeq \frac{\sum_{n=1}^\infty 2n(2n-1) c_{\scpt 2n} \phi^{2n-2}}{V_0}, \label{etacondition}
\end{align}
are violated if coefficients are too large. The primes in the above equations denote derivative with respect to $\phi$. The parameter region in the $\{c_2, c_4, c_6,...\}$-space where inflation can occur and generate the observed power spectrum is bounded by some $c^{\scpt\text{max}}_{\scpt 2n}$ for each $c_{\scpt 2n}$. As we will see below, $c^{\scpt\text{max}}_{\scpt 2n}$'s are determined by the energy scale parameter $V_0$. As we assume the probability distribution is uniform, the probability to have the inflation to occur around the energy scale $V_0$ is
\beq
P \propto \int^{\phi_{\rm ant}}_0 d\phi_i \int dV_0 \prod_{n=1}^\infty \left( \int_0^{c^{\scpt\text{max}}_{\scpt 2n}}d c_{\scpt 2n} \right) \propto  V_0 \phi_{\rm ant} \prod_{n=1}^{\infty} c^{\scpt\text{max}}_{\scpt 2n}. \label{PofV0}
\eeq
Note that we have simplified the problem by assuming $c^{\scpt\text{max}}_{\scpt 2n}$'s are independent on each other and a more detailed treatment will result in a probability slightly smaller than Eq.(\ref{PofV0}). Nevertheless, the main takeaway we can learn from such analysis will not be affected as we will explain below. Also note that we have included the probability distribution of the inflaton initial field value as advocated in Sec.\ref{subsec:InitialCondition}, and impose the anthropic constraint that inflation needs to last for more than $\calN_e^{\rm ant}$. 

\begin{figure}[tb]
\begin{center}
\includegraphics[width=\textwidth]{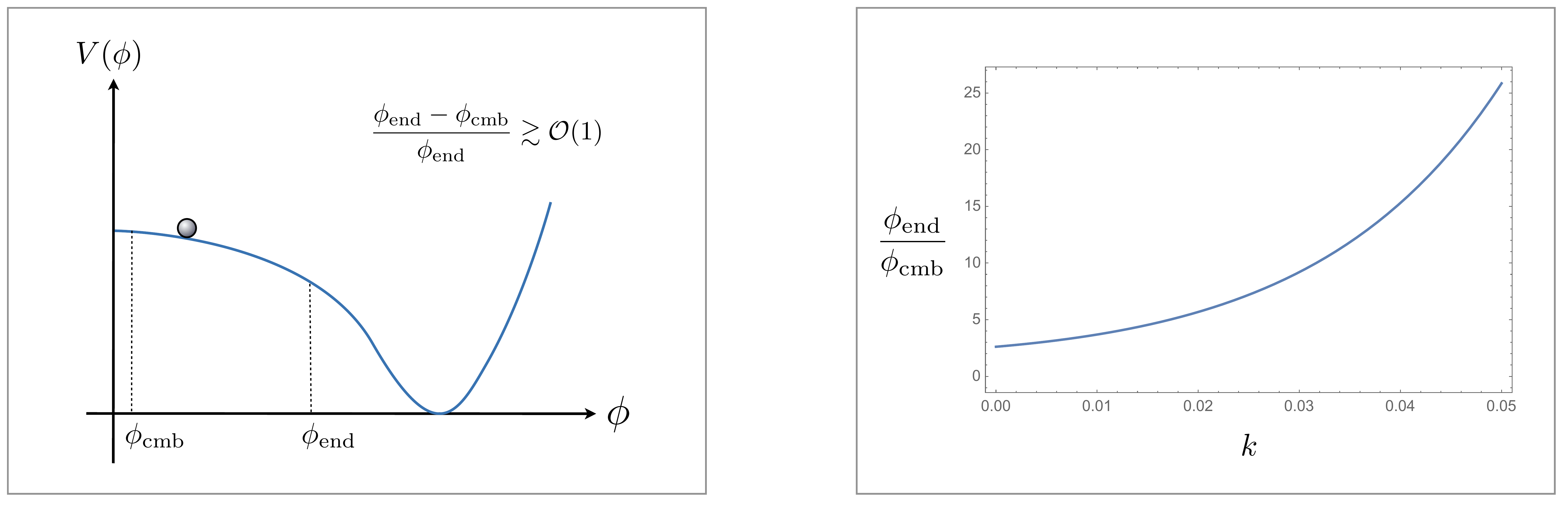}
\caption{For new inflation, the inflationary expansion occur mostly on the hilltop where $\phi$ is close to the origin. The inflation ends at $\phi_{\text{end}}$ where the potential violates the slow-roll condition and $\phi_\text{end} \gg \phi_\text{cmb}$.}
\label{fig:phiendphicmb}
\end{center}
\end{figure}

To find what $c^{\scpt\text{max}}_{\scpt 2n}$ is, we need to first know the field value $\phi_\text{end}$ when the inflation ends. This is determined by the number of e-folds $\calN^*_e$ between horizon exit and the end of the inflation, the inflation energy scale $V_0$, and the perturbation power spectrum $P_\zeta$. In particular, one has
\begin{align}
\calN_e = & \int H\,dt = \int \frac{H}{\dot{\phi}}\,d\phi = - \int \frac{V}{V'}\,d\phi = \int \frac{d\phi}{\sqrt{2\epsilon}}\simeq \frac{\phi_{\text{end}} - \phi_\text{cmb}}{\sqrt{2\epsilon}} = \frac{\Delta \phi}{\sqrt{2\epsilon}},\nonumber \\
 \Delta \phi \equiv & \phi_{\text{end}} - \phi_\text{cmb},
\end{align}
where we assumed the $\epsilon$ parameter to be nearly constant over the period of inflation. Its value can be determined by the perturbation power spectrum, 
\beq
\epsilon = \frac{1}{24\pi^2} \frac{V_0}{P_\zeta}.
\eeq
For potentials of a new inflation type, we typically have $\Delta \phi \simeq \phi_{\text{end}}$.
For example, for the potential
\beq
V= a- \frac{1}{2} b\, \phi^2 - c_n \phi^n,
\eeq
for which the relation between $\phi_{\text{end}}$ and $\phi_{\text{cmb}}$ is explicitly computed in Appendix \ref{Sec:Appendix}, the ratio $\phi_{\text{end}}/\phi_{\text{cmb}}$ depends only on the parameter $k\equiv b/a$ and is plotted in the right of Figure \ref{fig:phiendphicmb}. We see that the ratio grows for larger $k$, which is not surprising as a larger $k$ requires a smaller $\phi_{\text{cmb}}$ to maintain the same e-folds of inflation. 
As $\Delta \phi = \phi_{\text{end}} - \phi_\text{cmb}$ and $\phi_\text{end} \gtrsim \phi_\text{cmb}$, we actually have $\Delta \phi \simeq \phi_{\text{end}}$ and hance
\beq
\phi_{\scpt \text{end}} \simeq \frac{\calN^*_e}{\sqrt{12} \pi} \sqrt{\frac{V_0}{P_\zeta}}. \label{phiend}
\eeq
We define $c^{\epsilon}_{\scpt 2n}$ to be the value of $c_{\scpt 2n}$ such that the $c_{\scpt 2n} \phi^{2n}_{\scpt\text{end}}$ term alone in Eq.(\ref{epsiloncondition}) can violate the $\epsilon$ slow-roll condition, i.e.
\beq
\frac{1}{2} \left( \frac{2n c^{\epsilon}_{\scpt 2n} \phi^{2n-1}_{\scpt\text{end}}}{V_0} \right)^2 =1,
\eeq
which yields
\beq
c^{\epsilon}_{\scpt 2n} = \frac{V_0}{\sqrt{2}n} \left(\frac{{\calN^*_e}^2 V_0}{12\pi^2 P_\zeta} \right)^{\frac{1-2n}{2}}. \label{c2nepsilonmax}
\eeq
Similarly, we define $c^{\eta}_{\scpt 2n}$ such that the $c_{\scpt 2n} \phi^{2n}_{\scpt\text{end}}$ term alone in Eq.(\ref{etacondition}) can violate the $\eta$ slow-roll condition, which yields
\beq
c^{\eta}_{\scpt 2n} = \frac{V_0}{2n(2n-1)}\left( \frac{{\calN^*_e}^2 V_0}{12\pi^2 P_\zeta} \right)^{1-n}.
\label{c2netamax}
\eeq
We then define $c^{\scpt\text{max}}_{\scpt 2n}$ to be the minimum of $c^{\epsilon}_{\scpt 2n}$, $c^{\eta}_{\scpt 2n}$ and one,
\beq
c^{\scpt\text{max}}_{\scpt 2n} = \text{Min}\big[c^{\epsilon}_{\scpt 2n} , c^{\eta}_{\scpt 2n}, 1\big].
\eeq

Lastly, we estimate $\phi_{\rm ant}$ originating from scanning over the initial inflaton field value. From the total number of e-folds, we have
\beq
\calN^{tot}_e = - \int \frac{V}{V'}\,d\phi \simeq  \int \frac{V_0}{\sum 2n \, c_{2n}\, \phi^{2n-1}} d\phi \simeq 
\int \frac{V_0}{\sum 2n \, c^\eta_{2n}\, \phi^{2n-1}}\,d\phi \simeq \int \frac{V_0}{2n \, c^\eta_{2n}\, \phi^{2n-1}}\,d\phi
\eeq
where in the third equality we used $c^{\eta}_{2n}$ to replace $c_n$ because when the potential terms are relevant to the inflationary dynamics, their coefficients will be bounded by $c^{max}_{2n}$ and $c^{\eta}_{2n} < c^\epsilon_{2n}$  when the energy scale $V_0$ is small. In the last equality, we approximated the summation by the $n\mathchar`-$th term because all the relevant potential terms are comparable. After performing the integral from $\phi_i$ to $\phi_{end}$, because the integral is dominated by the $\phi_i$ term, we obtain
\beq
\calN^{tot}_e \simeq \frac{2n-1}{2(n-1)} \left(\frac{12 \pi^2 P_\zeta}{{\calN^*_e}^2 V_0}  \right)^{1-n} \phi_i^{2(1-n)},
\eeq
which gives
\beq
\phi_{\rm ant} \propto \sqrt{\frac{V_0}{P_\zeta}}.
\eeq

\begin{figure}[tb]
\begin{center}
\includegraphics[width=\textwidth]{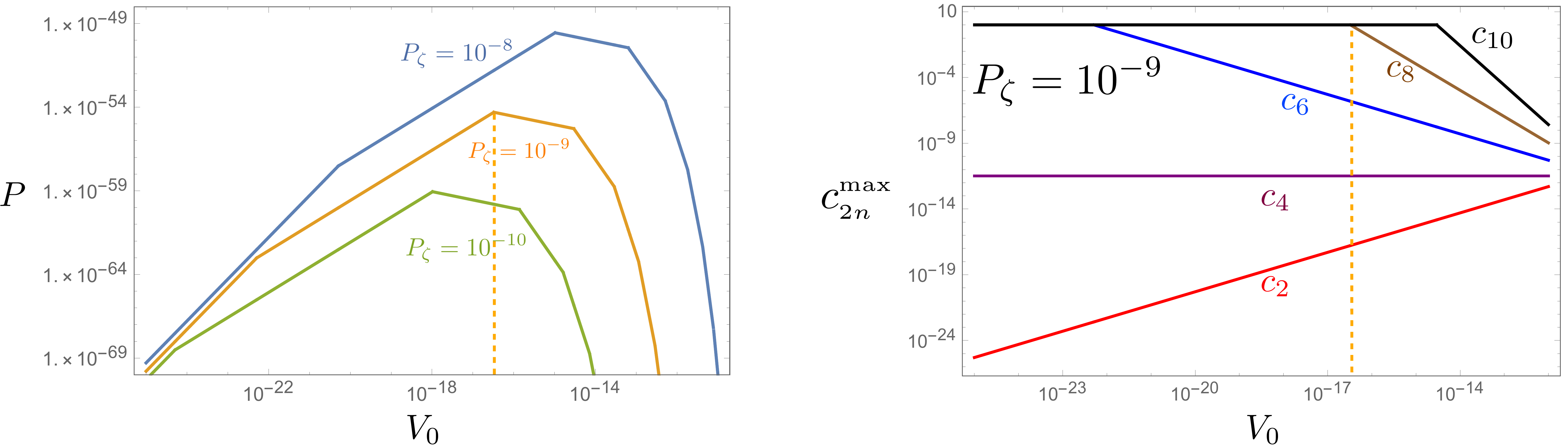}
\caption{\textbf{Left:} The probability $\calP$ for different primordial perturbation amplitude $P_\zeta$. \textbf{Right:} The values of $c^{\scpt \text{max}}_{\scpt 2}, c^{\scpt \text{max}}_{\scpt 4}, c^{\scpt \text{max}}_{\scpt 6}, c^{\scpt \text{max}}_{\scpt 8}$ and $c^{\scpt \text{max}}_{\scpt 10}$ as a function of $V_0$ with $P_\zeta = 10^{-9}$. Notice that $c^{\scpt \text{max}}_{\scpt 8}$ starts to decrease from 1 at $V_0 \sim 10^{-16}$, where $\calP$ reaches maximum. }
\label{fig:PV0diffPz}
\end{center}
\end{figure}

The probability $P \propto V_0 \phi_{\rm ant} \prod_{n=1}^{\infty} c^{\scpt\text{max}}_{\scpt 2n}$ is a function of $V_0$ and $P_\zeta$. We plot the unnormalized probability function in the left panel of Figure \ref{fig:PV0diffPz} for three different $P_\zeta$'s. One can see that the probability is strongly biased toward large perturbation. In addition, there are several kinks along each curves. To understand these kinks more,  it is illustrative to plot the first few $c^{\scpt \text{max}}_{\scpt 2n}$'s. In the right panel of Figure \ref{fig:PV0diffPz}, we see that the coefficients for higher dimensional operators, those with $n\geq 3$, have $c^{\scpt\text{max}}_{\scpt 2n} =1$ for small $V_0$. This is because $\phi_{\scpt\text{end}} \propto \sqrt{V_0}$ as shown in Eq.(\ref{phiend}) and hence for small $V_0$, the higher-dimensional operators are Planck-suppressed and irrelevant to inflationary dynamics. This is also the reason why assuming $c^{\scpt\text{max}}_{\scpt 2n}$'s are independent on each other does not change the qualitative result. Only the lower dimensional coefficients can affect the higher ones but not vice versa.  As we increase the inflation energy scale, the field displacement becomes larger and hence higher-dimensional operators become relevant and their required coefficients start to decrease from one. This translates to the kinks shown in the left panel of Figure \ref{fig:PV0diffPz}. For instance,  in Figure \ref{fig:PV0diffPz} we see that, for $P_\zeta = 10^{-9}$, $V_0 \sim 10^{-17}$ indicated by the orange dashed line is precisely the scale where the octet term starts to be relevant and require fine-tuning.  Also note that, regardless of the value of  $P_\zeta$, the amount of fine-tuning is minimal when the octet operator just became relevant.

We compute the probability to obtain a cosmic perturbation $P_\zeta$. As we have observed, the probability $P$ peaks at the point when the octet becomes relevant, i.e.~when $c^\eta_{8} =1$. This gives us the energy scale $V^{max}_0$ where the fine-tuning is minimal, 
\beq
V^{max}_0 = \frac{1}{2\sqrt{14}} \left( \frac{{\calN^*_e}^2}{12\pi^2 P_\zeta} \right)^{-\frac{3}{2}}.
\eeq
When we integrate out $V_0$ the integration is dominated by the region around $V^{max}_0$, we therefore have
\beq
P \propto V^{max}_0 \phi_{\rm ant} c^{max}_{2} c^{max}_{4} c^{max}_{6} \propto P_\zeta^{\frac{19}{4}}.
\eeq
This can be understood as $P_\zeta$ obeying the distribution
\begin{align}
P(P_\zeta) dP_\zeta \propto P_\zeta^{\frac{15}{4}} dP_\zeta.
\end{align}
The probability is strongly biased toward large $P_\zeta$. Unless there exists a strong anthropic bound disfavoring $P_\zeta$ larger than the observed one, it is unlikely that general new inflation with $Z_2$ symmetry results in our observed universe. Our analysis is also applicable to the case with $U(1)$ symmetry because the radial direction is essentially $Z_2$ symmetric, while the angular direction is flat and does not affect the inflationary dynamics at the background level.

For completeness, we continue our further analysis of general new inflation with $Z_2$ symmetry in the next section. In particular, assuming that the perturbation amplitude $P_\zeta$ is fixed to the observed value for some reason, we investigate the probability distribution of spectral index $n_s$. We will find that it is probabilistically favored to have a spectral index $n_s\simeq 0.96$, which is quite remarkable.

\section{New Inflation and the Most Probable Spectral Index}\label{Sec:Appendix}
In Appendix \ref{Sec:FineTuneGeneral} we found that when considering inflation with $Z_2$ symmetry, we need to fine-tune terms at least up to the octet order,
\beq
V = V_0 + c_{\scpt 2} \phi^2 + c_{\scpt 4} \phi^4 + c_{\scpt 6} \phi^6 + c_{\scpt 8} \phi^8. \label{V8}
\eeq
Nevertheless, for simplicity we will consider potential of the form
\beq
V= a- \frac{1}{2} b\, \phi^2 - c_n \phi^n - c_m \phi^m  \label{nmModel}
\eeq
where the $c_n$ term dominates over the $c_m$ term and the latter is treated perturbatively. Namely, we consider the case where $c_m$ is sufficiently small and $c_m \phi^m < c_n \phi^n$. We can then extract the physics of the complete octet model, Eq.(\ref{V8}), by extrapolation. In Eq.(\ref{nmModel}) we make quadratic term explicitly negative as we now consider cases where the inflaton rolls down from $\phi=0$.

\begin{figure}[tb]
\begin{center}
\includegraphics[width=\textwidth]{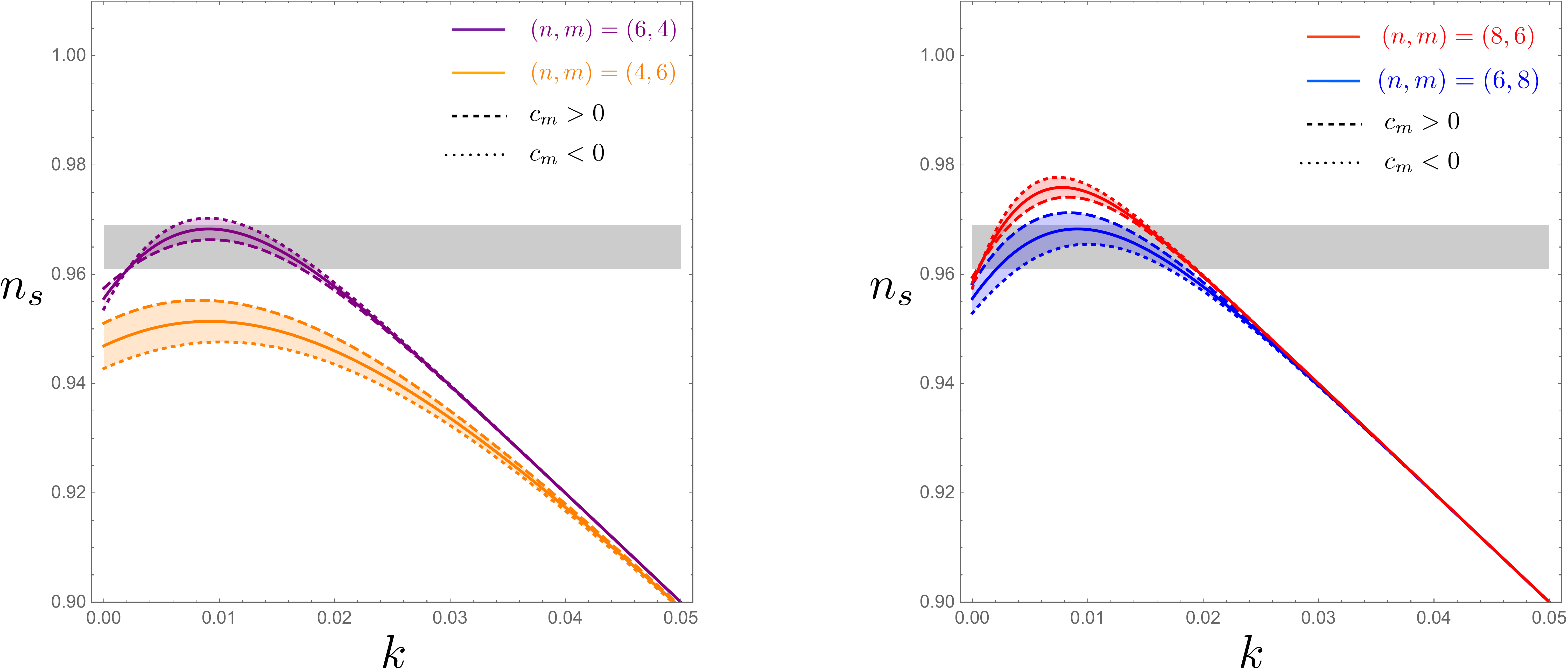}
\caption{The $n_s - k$ plot for different set of $(n ,m)$. The dashed lines corresponds to the boundary of positive  perturbation, $c_m>0$, while the dotted ones represents negative perturbation, $c_m<0$. The grey region shows the current observational value. }
\label{fig:nm6864}
\end{center}
\end{figure}

The number of e-folds $\calN_e$ which the inflation would last before it ends and the corresponding field value $\phi_{\scpt\calN_e}$ has the relation
\begin{align}
\calN_e & =  \int^{\phiNe}_{\phi_{end}} \frac{V}{ \p V /\p \phi}\,d\phi \nonumber \\
        & \simeq \int^{\phiNe}_{\phi_{end}} \frac{a}{ - b\phi - n c_n \phi^{n-1} - m c_m \phi^{m-1}} \,d\phi   \nonumber \\
        & \simeq  a \int^{\phiNe}_{\phi_{end}} \left[\frac{1}{b \phi + n c_n \phi^{n-1}} - \frac{m \phi^{m-1}}{b \phi + n c_n \phi^{n-1}} \,c_m  \right] d \phi  \nonumber \\
        & = \frac{1}{(n-2) k } \ln \left( \frac{k \phiNe^{2-n} + \frac{c_n}{a} n}{k \phi^{2-n}_{end} + \frac{c_n}{a} n} \right)
        	 	- \frac{1}{k} \left[ \calF(\phi_{end}) - \calF({\phiNe})  \right] c_m  \label{nmNephi}
\end{align}
where we defined $k\equiv  b/a$ and 
\beq
\calF(\phi) \equiv \frac{m}{n-2} \frac{\phi^{m-2}}{a k} \left\{ \left(\frac{n-m}{m-2} \right)  {}_2F_1 \left(1, \frac{m-2}{n-2}; \frac{n+m-4}{n-2}; - \frac{n c_n}{a k} \phi^{n-2}\right)  + \frac{a k}{a k+ n c_n \phi^{n-2}}  \right\}.
\eeq
Here ${_2F_1}$ is the hypergeometric function and $\phi_{end}$ is the field value where the inflation ends, determined by the slow-roll condition. In particular, the inflation ends when the $\eta$-parameter reaches -1, which yields
\beq
\phi^{n-2}_{end} \simeq \frac{a}{n(n-1) c_n}.
\eeq
Solving Eq.(\ref{nmNephi}) for $\phiNe(\calN_e)$ perturbatively in $c_m$, that is, with $\phiNe(\calN_e) = \phiNe^{(0)} + c_m\, \phiNe^{(1)}$, one has
\beq
\phiNe^{(0)} = \left[ \frac{a}{c_n} k f_n (k)\right]^{\frac{1}{n-2}}, \quad\text{ and } \quad \phiNe^{(1)} = \phi_0 \, c_n^{-\frac{m-2}{n-2}} a^{\frac{m-n}{n-2}} g_{n,m}(k) \label{phiNe}
\eeq
where
\beq
f_n(k) \equiv \frac{1}{n} \frac{1}{\left( \left[1+(n-1) k \right] e^{(n-2) k \calN_e} -1 \right)} \label{fn}
\eeq
and
\begin{align}
g_{n,m}(k)  \equiv & \, \frac{m (n-m)}{(n-2)(m-2)}  \left(1 +  n f_n \right) \times \nonumber \\
& \hspace{1cm} \Bigg\{
k^\frac{m-n}{n-2} f_n^{\frac{m-2}{n-2}} \left[  {}_2F_1\left(1, \frac{m-2}{n-2}; \frac{n+m-4}{n-2} ,-n f_n\right) + \frac{1}{1+ n f_n} \right]  \nonumber \\
&\hspace{2cm} \left. - \left(\frac{1}{n(n-1)} \right)^\frac{m-2}{n-2} \frac{1}{k} \left[ {}_2F_1\left(1, \frac{m-2}{n-2}; \frac{n+m-4}{n-2}, -\frac{1}{(n-1) k}\right) + \frac{1}{1+ \frac{1}{(n-1)k}} \right]
\right\}.
\end{align}
The spectral index $n_s$ and perturbation power spectrum $P_\zeta$ when inflation can last another $\calN_e$ of e-folds before it ends is then given by
\begin{align}
n_s & = 1 + 2 \eta - 6 \epsilon \nonumber \\
       & = 1- 2k - 2n(n-1) k f_n - 3 a^\frac{2}{n-2} c_n^{-\frac{2}{n-2}} k^\frac{2n-2}{n-2} \left(1+ n f_n \right)^2 f_n^{\frac{2}{n-2}} \nonumber \\
       & \hspace{1cm} + \Bigg\{ 2 a^\frac{m-n}{n-2} c_n^{-\frac{m-2}{n-2}} \left[m (1-m) k^\frac{m-2}{n-2} f_n^\frac{m-2}{n-2} - n(n-1)(n-2) k \, f_n \, g_{n,m}\right] \nonumber \\
       & \hspace{1.5cm} - 6 a^\frac{m-n+2}{n-2} c_n^{-\frac{m}{n-2}} k^{\frac{n}{n-2}}f_n^\frac{2}{n-2} (1+ n f_n)
       		\left( m k^\frac{m-2}{n-2} f_n^\frac{m-2}{n-2} + k\, g_{n,m} + n(n-1) k \,f_n \, g_{n,m} \right) \Bigg\} c_m,
		\label{nmns}
\end{align}
\begin{align}
P_\zeta & = \frac{1}{24\pi^2} \frac{V}{\epsilon} \nonumber \\
	     & = \frac{ a^\frac{n-4}{n-2} k^\frac{-2n+2}{n-2} f_n^{-\frac{2}{n-2}} }{12\pi^2 \left(1+n f_n \right)^2} 
	     		\left[c^\frac{2}{n-2}_n - c^\frac{4-m}{n-2}_n \, c_m \, \frac{2 a^\frac{m-n}{n-2}\left( m k^\frac{m-n}{n-2} f^\frac{m-2}{n-2}_n + g_{n,m} + n(n-1) \, f_n \,g_{n,m} \right) }{(1+ n f_n)}  \right]. \label{nmPzeta}
\end{align}
By taking the inverse of Eq.(\ref{nmPzeta}) perturbatively in $c_m$, one obtain $c_n = c_n^{(0)} + c_m \,c_n^{(1)}$ with
\beq
c^{(0)}_n = (12\pi^2)^\frac{n-2}{2} a^\frac{4-n}{2} k^{n-1} f_n \left(1+n f_n \right)^{n-2} P_\zeta^\frac{n-2}{2} , \label{nmcn0}
\eeq
and
\beq
c^{(1)}_n = \frac{n-2}{2} {c^{(0)}_n}^\frac{n-m+2}{n-2} \frac{2 a^\frac{m-n}{n-2} \left( m k^\frac{m-n}{n-2} f_n^\frac{m-2}{n-2} + g_{n,m} + n(n-1) f_n \, g_{n,m} \right)}{1+ n f_n}. \label{nmcn1}
\eeq
With Eq.(\ref{nmns}), (\ref{nmcn0}) and (\ref{nmcn1}), the relation between the spectral index $n_s$ and the parameter $k$ is plotted in Figure \ref{fig:nm6864} for various set of $(n,m)$. We set the parameter $a=10^{-16}$, the number of e-folds $\calN_e = 55$, and the perturbation power spectrum to the observed value, $P_\zeta=2.2 \times 10^{-9}$. The parameter $c_m$ is bounded by the requirement that the perturbation holds, i.e. $c_m \phi^m < c_n \phi^n$. As $\phi_{end} > \phiNe$, the bound of $c_m$ is therefore
\beq
|c_m| < 
\begin{cases}
c_n \phiNe^{n-m} & \quad  ( n>m ) , \\
\\
c_n \phi _{end}^{n-m} & \quad (n<m).
\end{cases}
\eeq

\begin{figure}[tb]
\begin{center}
\includegraphics[width=\textwidth]{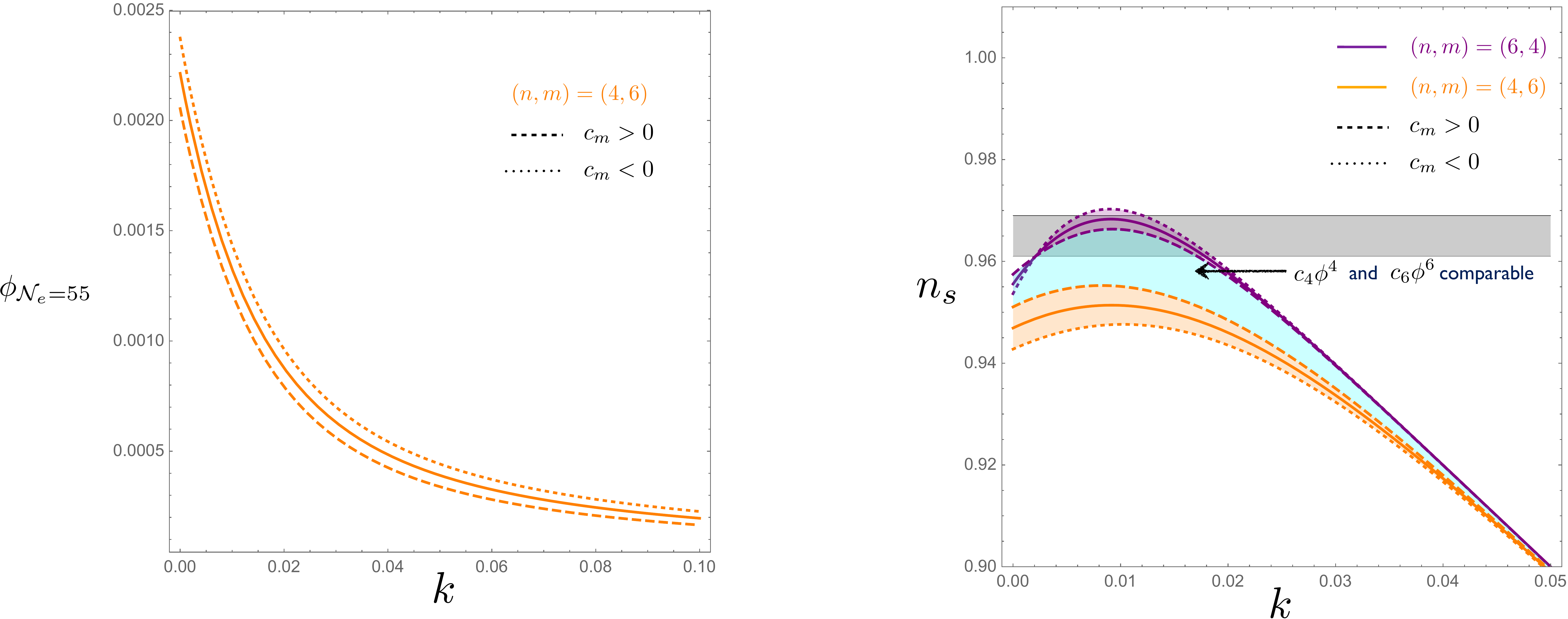}
\caption{(Left) The $\phi - k$ plot for $(n, m) = (4, 6)$. One can see that for the same $k$, $\phi$ is larger when $c_m<0$. (Right) Even though our treatment is perturbative, the case where $c_n \phi^n$ and $c_m \phi^m$ are comparable should continuously connect the two perturbative regions.}
\label{fig:phik46}
\end{center}
\end{figure}

We first look at the case with quartic and sextic terms, shown on the left of Figure \ref{fig:nm6864}. Note that the quartic-dominant case is excluded by observation. One generic feature that appears for all $(n,m)$ is that when $n<m$, a positive perturbation ($c_m>0$) leads to a spectral index closer to scale-invariance, i.e. $n_s=1$, while a negative perturbation ($c_m<0$) makes $n_s$ deviate away from $1$. This might be counterintuitive as the potential
\beq
V= a- \frac{1}{2} b \phi^2 - c_n \phi^n - c_m \phi^m,
\eeq
is flatter when $c_m<0$ and we would have expected a spectral index closer to $1$. However, a flatter potential also means the inflaton moves slower before it reaches $\phi_{end}$, and hence the field value can be closer to $\phi_{end}$ (but farther away from zero) while giving the same number of e-foldings as shown in the left of Figure \ref{fig:phik46}. As the field is farther away from zero where the potential is the flattest, the spectral index can deviate from -1. The two effects, flatter potential and larger $\phiNe$, compete with each other and the latter wins when $n<m$. On the other hand, for $n>m$ the effect of flatter potential dominates and a negative perturbation ($c_m <0$) leads to spectral index closer to 1. The fact that the two cases, $n>m$ and $n<m$, behave oppositely and the region of positive perturbation lies between two unperturbed curves makes us confident that one can extrapolate our perturbative treatment to the case where the $c_n$ term and $c_m$ term comparable -- it simply lies between the two perturbative regions where one dominates the other, as shown in the right in Figure \ref{fig:phik46}.

\begin{figure}[tb]
\begin{center}
\includegraphics[width=0.4\textwidth]{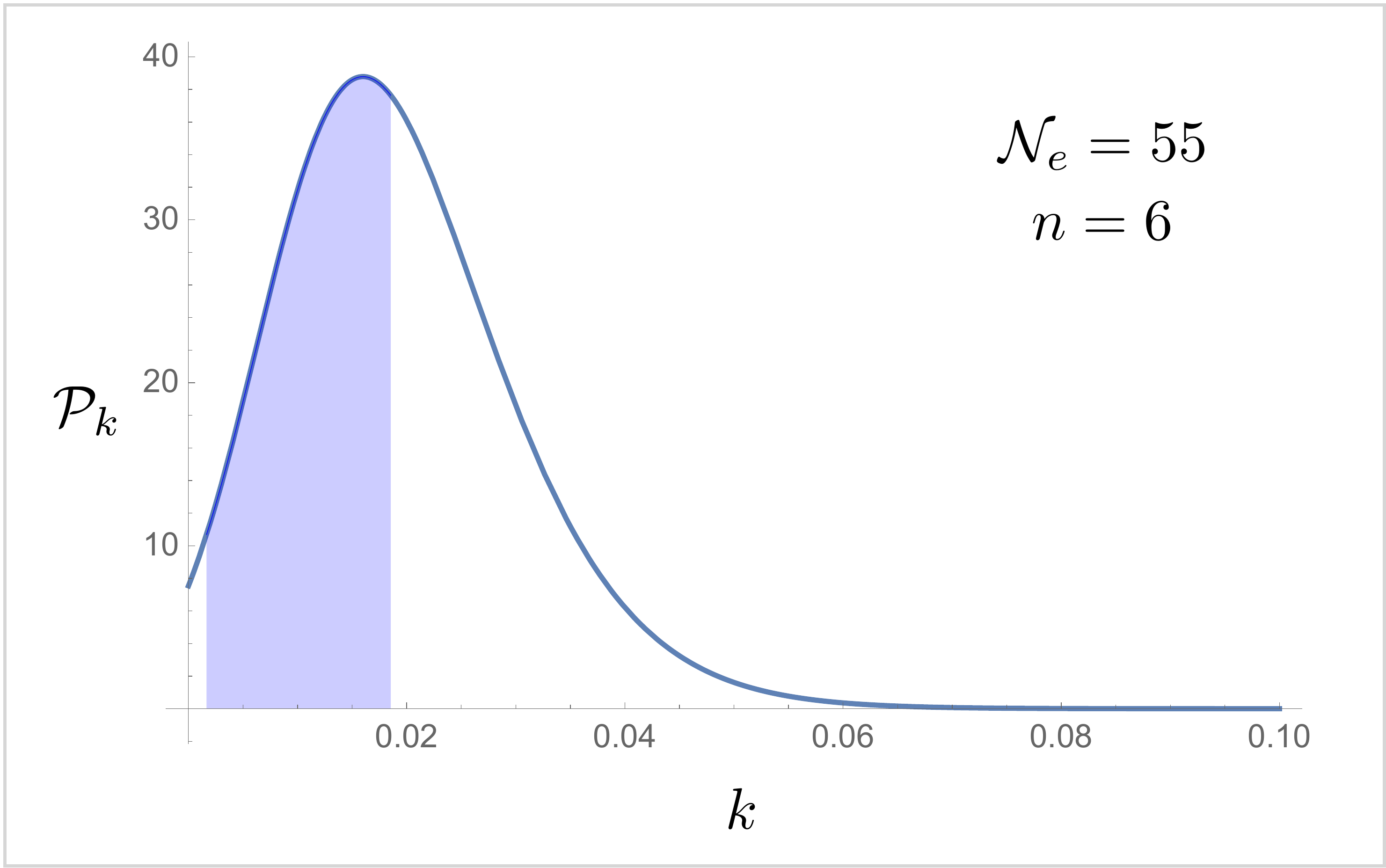}
\caption{The normalized probability distribution of $k$ for $n=6$. The shaded area corresponds to the interval of $k$ that yields spectral index $n_s>0.96$.}
\label{fig:Pk6}
\end{center}
\end{figure}

In Figure \ref{fig:nm6864} we see that sextic dominant models, $n=6$, with either quartic perturbation ($m=4$) or octet perturbation ($m=8$), fit the observation quite well when $k\lesssim 0.018$. We discuss the chance for $k$ to lie in this region.
We will focus on the following analysis without perturbation,
\beq
V= a- \frac{1}{2} b\, \phi^2 - c_n \phi^n
\eeq
as additional perturbation $c_m \phi^m$ does not change the end result significantly. We assume the parameter in the Lagrangian $a$, $b$ and $c_n$ has uniform probability distribution and also take the probabilistic nature of the initial condition into account.  Using the definition $b = a k$, Eq.(\ref{phiNe}) and Eq.(\ref{nmcn0}), one has
\begin{align}
\int^{\phi_{\rm ant}}_0 d\phi_i\int da\,db\,dc_n 
=\int da\,dP_\zeta \,dk \,(12\pi^2)^{\scpt \frac{n-3}{2}} a^{\scpt\frac{7-n}{2}} P_\zeta^{\scpt\frac{n-5}{2}} k^{n-2} f_n \left(1+ n\, f_n \right)^{n-3}, \label{measureNonSUSY}
\end{align}
where we need to integrate over $a$ to obtain the probability distribution of $k$ for a given $P_\zeta$. For $n=6$ and 8, the integration over $a$ is divergent with an upper bound $a_{max}$ that is around $10^{-17}$ where only terms below the octec order require fine-tuning. The exact upper bound $a_{max}$ is correlated with the upper bound for $c_n$, but most importantly $a_{max}$ is independent on $k$ and hence the integration over $a$ does not give an additional $k$-dependence. In sum, for the cases of interest, the probability distribution of $k$ is
\beq
 \calP_k \,dk =  k^{n-2} f_n \left(1+ n\, f_n \right)^{n-3} \,dk, \label{Pk}
\eeq
where $f_n(k)$ is defined in Eq.(\ref{fn}), and the plot for $n=6$ is given in Figure \ref{fig:Pk6}. The shaded area corresponds to the interval $\calI$ of $k$ that yields spectral index $n_s >0.96$. The probability for $k$ to lie in this region for $n=6$ is
\beq
P_{n_s>0.96} = \frac{\int_{\calI} \calP_k\,dk}{\int^\infty_0 \calP_k\,dk} \simeq 0.49.
\eeq
and the distribution $\calP_k$ peaks at $k\simeq 0.016$, which yields a spectral index of $n_s = 0.963$. Overall it is quite remarkable that once one matches the observed perturbation power spectrum $P_\zeta$, there is about a few ten chance to achieve the observed spectral index without much further fine-tuning in general new inflation with $Z_2$ symmetry. But as we discussed in Sec.\ref{Sec:FineTuneGeneral}, the observed $P_\zeta$ can be obtained without significant fine-tuning only if there is a strong anthropic bound on $P_\zeta$ right at the observed value which seems to be unlikely.

\addcontentsline{toc}{section}{References}
\bibliographystyle{utphys}
\bibliography{RefsInflation}

\providecommand{\href}[2]{#2}\begingroup\raggedright\begin{thebibliography}{10}

\bibitem{Guth:1980zm}
A.~H. Guth, ``{The Inflationary Universe: A Possible Solution to the Horizon
  and Flatness Problems},''
\href{http://dx.doi.org/10.1103/PhysRevD.23.347}{{\em Phys. Rev.} {\bfseries
  D23} (1981) 347--356}.

\bibitem{Kazanas:1980tx}
D.~Kazanas, ``{Dynamics of the Universe and Spontaneous Symmetry Breaking},''
\href{http://dx.doi.org/10.1086/183361}{{\em Astrophys. J.} {\bfseries 241}
  (1980) L59--L63}.

\bibitem{Mukhanov:1981xt}
V.~F. Mukhanov and G.~V. Chibisov, ``{Quantum Fluctuations and a Nonsingular
  Universe},'' {\em JETP Lett.} {\bfseries 33} (1981) 532--535.
[Pisma Zh. Eksp. Teor. Fiz.33,549(1981)].

\bibitem{Hawking:1982cz}
S.~W. Hawking, ``{The Development of Irregularities in a Single Bubble
  Inflationary Universe},''
\href{http://dx.doi.org/10.1016/0370-2693(82)90373-2}{{\em Phys. Lett.}
  {\bfseries 115B} (1982) 295}.

\bibitem{Starobinsky:1982ee}
A.~A. Starobinsky, ``{Dynamics of Phase Transition in the New Inflationary
  Universe Scenario and Generation of Perturbations},''
\href{http://dx.doi.org/10.1016/0370-2693(82)90541-X}{{\em Phys. Lett.}
  {\bfseries 117B} (1982) 175--178}.

\bibitem{Guth:1982ec}
A.~H. Guth and S.~Y. Pi, ``{Fluctuations in the New Inflationary Universe},''
\href{http://dx.doi.org/10.1103/PhysRevLett.49.1110}{{\em Phys. Rev. Lett.}
  {\bfseries 49} (1982) 1110--1113}.

\bibitem{Bardeen:1983qw}
J.~M. Bardeen, P.~J. Steinhardt, and M.~S. Turner, ``{Spontaneous Creation of
  Almost Scale - Free Density Perturbations in an Inflationary Universe},''
\href{http://dx.doi.org/10.1103/PhysRevD.28.679}{{\em Phys. Rev.} {\bfseries
  D28} (1983) 679}.

\bibitem{Linde:1981mu}
A.~D. Linde, ``{A New Inflationary Universe Scenario: A Possible Solution of
  the Horizon, Flatness, Homogeneity, Isotropy and Primordial Monopole
  Problems},''
\href{http://dx.doi.org/10.1016/0370-2693(82)91219-9}{{\em Phys. Lett.}
  {\bfseries 108B} (1982) 389--393}.

\bibitem{Albrecht:1982wi}
A.~Albrecht and P.~J. Steinhardt, ``{Cosmology for Grand Unified Theories with
  Radiatively Induced Symmetry Breaking},''
\href{http://dx.doi.org/10.1103/PhysRevLett.48.1220}{{\em Phys. Rev. Lett.}
  {\bfseries 48} (1982) 1220--1223}.

\bibitem{Starobinsky:1980te}
A.~A. Starobinsky, ``{A New Type of Isotropic Cosmological Models Without
  Singularity},''
\href{http://dx.doi.org/10.1016/0370-2693(80)90670-X}{{\em Phys. Lett.}
  {\bfseries B91} (1980) 99--102}.

\bibitem{Susskind:2003kw}
L.~Susskind, ``{The Anthropic landscape of string theory},''
\href{http://arxiv.org/abs/hep-th/0302219}{{\ttfamily arXiv:hep-th/0302219
  [hep-th]}}.

\bibitem{Bousso:2000xa}
R.~Bousso and J.~Polchinski, ``{Quantization of four form fluxes and dynamical
  neutralization of the cosmological constant},''
  \href{http://dx.doi.org/10.1088/1126-6708/2000/06/006}{{\em JHEP} {\bfseries
  06} (2000) 006},
\href{http://arxiv.org/abs/hep-th/0004134}{{\ttfamily arXiv:hep-th/0004134
  [hep-th]}}.

\bibitem{Vilenkin:1994ua}
A.~Vilenkin, ``{Predictions from quantum cosmology},''
  \href{http://dx.doi.org/10.1103/PhysRevLett.74.846}{{\em Phys. Rev. Lett.}
  {\bfseries 74} (1995) 846--849},
\href{http://arxiv.org/abs/gr-qc/9406010}{{\ttfamily arXiv:gr-qc/9406010
  [gr-qc]}}.

\bibitem{Weinberg:1987dv}
S.~Weinberg, ``{Anthropic Bound on the Cosmological Constant},''
\href{http://dx.doi.org/10.1103/PhysRevLett.59.2607}{{\em Phys. Rev. Lett.}
  {\bfseries 59} (1987) 2607}.

\bibitem{Freivogel:2005vv}
B.~Freivogel, M.~Kleban, M.~Rodriguez~Martinez, and L.~Susskind,
  ``{Observational consequences of a landscape},''
  \href{http://dx.doi.org/10.1088/1126-6708/2006/03/039}{{\em JHEP} {\bfseries
  03} (2006) 039},
\href{http://arxiv.org/abs/hep-th/0505232}{{\ttfamily arXiv:hep-th/0505232
  [hep-th]}}.

\bibitem{Guth:2013sya}
A.~H. Guth, D.~I. Kaiser, and Y.~Nomura, ``{Inflationary paradigm after Planck
  2013},'' \href{http://dx.doi.org/10.1016/j.physletb.2014.03.020}{{\em Phys.
  Lett.} {\bfseries B733} (2014) 112--119},
\href{http://arxiv.org/abs/1312.7619}{{\ttfamily arXiv:1312.7619
  [astro-ph.CO]}}.

\bibitem{Coleman:1980aw}
S.~R. Coleman and F.~De~Luccia, ``{Gravitational Effects on and of Vacuum
  Decay},''
\href{http://dx.doi.org/10.1103/PhysRevD.21.3305}{{\em Phys. Rev.} {\bfseries
  D21} (1980) 3305}.

\bibitem{Gott:1982zf}
J.~R. Gott, ``{Creation of Open Universes from de Sitter Space},''
\href{http://dx.doi.org/10.1038/295304a0}{{\em Nature} {\bfseries 295} (1982)
  304--307}.

\bibitem{Linde:1995rv}
A.~D. Linde and A.~Mezhlumian, ``{Inflation with $\Omega \neq 1$},''
  \href{http://dx.doi.org/10.1103/PhysRevD.52.6789}{{\em Phys. Rev.} {\bfseries
  D52} (1995) 6789--6804},
\href{http://arxiv.org/abs/astro-ph/9506017}{{\ttfamily arXiv:astro-ph/9506017
  [astro-ph]}}.

\bibitem{Vilenkin:1996ar}
A.~Vilenkin and S.~Winitzki, ``{Probability distribution for omega in open
  universe inflation},'' \href{http://dx.doi.org/10.1103/PhysRevD.55.548}{{\em
  Phys. Rev.} {\bfseries D55} (1997) 548--559},
\href{http://arxiv.org/abs/astro-ph/9605191}{{\ttfamily arXiv:astro-ph/9605191
  [astro-ph]}}.

\bibitem{Aghanim:2018eyx}
{\bfseries Planck} Collaboration, N.~Aghanim {\em et~al.}, ``{Planck 2018
  results. VI. Cosmological parameters},''
\href{http://arxiv.org/abs/1807.06209}{{\ttfamily arXiv:1807.06209
  [astro-ph.CO]}}.

\bibitem{Martel:1997vi}
H.~Martel, P.~R. Shapiro, and S.~Weinberg, ``{Likely values of the cosmological
  constant},'' \href{http://dx.doi.org/10.1086/305016}{{\em Astrophys. J.}
  {\bfseries 492} (1998) 29},
\href{http://arxiv.org/abs/astro-ph/9701099}{{\ttfamily arXiv:astro-ph/9701099
  [astro-ph]}}.

\bibitem{Barrow:1993}
J.~D. {Barrow} and P.~{Saich}, ``{Growth of large-scale structure with a
  cosmological constant},''
  \href{http://dx.doi.org/10.1093/mnras/262.3.717}{{\em Monthly Notices of the
  Royal Astronomical Society} {\bfseries 262} (June, 1993) 717--725}.

\bibitem{Barrow:1988yia}
J.~D. Barrow and F.~J. Tipler, {\em {The Anthropic Cosmological Principle}}.
\newblock Oxford U. Pr., Oxford,
1986.
\newblock

\bibitem{Bousso:2010vi}
R.~Bousso and R.~Harnik, ``{The Entropic Landscape},''
  \href{http://dx.doi.org/10.1103/PhysRevD.82.123523}{{\em Phys. Rev.}
  {\bfseries D82} (2010) 123523},
\href{http://arxiv.org/abs/1001.1155}{{\ttfamily arXiv:1001.1155 [hep-th]}}.

\bibitem{Tegmark:1997in}
M.~Tegmark and M.~J. Rees, ``{Why is the Cosmic Microwave Background
  fluctuation level 10**(-5)?},'' \href{http://dx.doi.org/10.1086/305673}{{\em
  Astrophys. J.} {\bfseries 499} (1998) 526--532},
\href{http://arxiv.org/abs/astro-ph/9709058}{{\ttfamily arXiv:astro-ph/9709058
  [astro-ph]}}.

\bibitem{Tegmark:2005dy}
M.~Tegmark, A.~Aguirre, M.~Rees, and F.~Wilczek, ``{Dimensionless constants,
  cosmology and other dark matters},''
  \href{http://dx.doi.org/10.1103/PhysRevD.73.023505}{{\em Phys. Rev.}
  {\bfseries D73} (2006) 023505},
\href{http://arxiv.org/abs/astro-ph/0511774}{{\ttfamily arXiv:astro-ph/0511774
  [astro-ph]}}.

\bibitem{Linde:1993xx}
A.~D. Linde, D.~A. Linde, and A.~Mezhlumian, ``{From the Big Bang theory to the
  theory of a stationary universe},''
  \href{http://dx.doi.org/10.1103/PhysRevD.49.1783}{{\em Phys. Rev.} {\bfseries
  D49} (1994) 1783--1826},
\href{http://arxiv.org/abs/gr-qc/9306035}{{\ttfamily arXiv:gr-qc/9306035
  [gr-qc]}}.

\bibitem{Bousso:2006ev}
R.~Bousso, ``{Holographic probabilities in eternal inflation},''
  \href{http://dx.doi.org/10.1103/PhysRevLett.97.191302}{{\em Phys. Rev. Lett.}
  {\bfseries 97} (2006) 191302},
\href{http://arxiv.org/abs/hep-th/0605263}{{\ttfamily arXiv:hep-th/0605263
  [hep-th]}}.

\bibitem{DeSimone:2008bq}
A.~De~Simone, A.~H. Guth, M.~P. Salem, and A.~Vilenkin, ``{Predicting the
  cosmological constant with the scale-factor cutoff measure},''
  \href{http://dx.doi.org/10.1103/PhysRevD.78.063520}{{\em Phys. Rev.}
  {\bfseries D78} (2008) 063520},
\href{http://arxiv.org/abs/0805.2173}{{\ttfamily arXiv:0805.2173 [hep-th]}}.

\bibitem{Bousso:2008hz}
R.~Bousso, B.~Freivogel, and I.-S. Yang, ``{Properties of the scale factor
  measure},'' \href{http://dx.doi.org/10.1103/PhysRevD.79.063513}{{\em Phys.
  Rev.} {\bfseries D79} (2009) 063513},
\href{http://arxiv.org/abs/0808.3770}{{\ttfamily arXiv:0808.3770 [hep-th]}}.

\bibitem{Nomura:2011dt}
Y.~Nomura, ``{Physical Theories, Eternal Inflation, and Quantum Universe},''
  \href{http://dx.doi.org/10.1007/JHEP11(2011)063}{{\em JHEP} {\bfseries 11}
  (2011) 063},
\href{http://arxiv.org/abs/1104.2324}{{\ttfamily arXiv:1104.2324 [hep-th]}}.

\bibitem{Garriga:2012bc}
J.~Garriga and A.~Vilenkin, ``{Watchers of the multiverse},''
  \href{http://dx.doi.org/10.1088/1475-7516/2013/05/037}{{\em JCAP} {\bfseries
  1305} (2013) 037},
\href{http://arxiv.org/abs/1210.7540}{{\ttfamily arXiv:1210.7540 [hep-th]}}.

\bibitem{Ovrut:1983my}
B.~A. Ovrut and P.~J. Steinhardt, ``{Supersymmetry and Inflation: A New
  Approach},''
\href{http://dx.doi.org/10.1016/0370-2693(83)90551-8}{{\em Phys. Lett.}
  {\bfseries 133B} (1983) 161--168}.

\bibitem{Holman:1984yj}
R.~Holman, P.~Ramond, and G.~G. Ross, ``{Supersymmetric Inflationary
  Cosmology},''
\href{http://dx.doi.org/10.1016/0370-2693(84)91729-5}{{\em Phys. Lett.}
  {\bfseries 137B} (1984) 343--347}.

\bibitem{Goncharov:1983mw}
A.~B. Goncharov and A.~D. Linde, ``{Chaotic Inflation in Supergravity},''
\href{http://dx.doi.org/10.1016/0370-2693(84)90027-3}{{\em Phys. Lett.}
  {\bfseries 139B} (1984) 27--30}.

\bibitem{Coughlan:1984yk}
G.~D. Coughlan, R.~Holman, P.~Ramond, and G.~G. Ross, ``{Supersymmetry and the
  Entropy Crisis},''
\href{http://dx.doi.org/10.1016/0370-2693(84)91043-8}{{\em Phys. Lett.}
  {\bfseries 140B} (1984) 44--48}.

\bibitem{Copeland:1994vg}
E.~J. Copeland, A.~R. Liddle, D.~H. Lyth, E.~D. Stewart, and D.~Wands, ``{False
  vacuum inflation with Einstein gravity},''
  \href{http://dx.doi.org/10.1103/PhysRevD.49.6410}{{\em Phys. Rev.} {\bfseries
  D49} (1994) 6410--6433},
\href{http://arxiv.org/abs/astro-ph/9401011}{{\ttfamily arXiv:astro-ph/9401011
  [astro-ph]}}.

\bibitem{Tegmark:2004qd}
M.~Tegmark, ``{What does inflation really predict?},''
  \href{http://dx.doi.org/10.1088/1475-7516/2005/04/001}{{\em JCAP} {\bfseries
  0504} (2005) 001},
\href{http://arxiv.org/abs/astro-ph/0410281}{{\ttfamily arXiv:astro-ph/0410281
  [astro-ph]}}.

\bibitem{Masoumi:2016eag}
A.~Masoumi, A.~Vilenkin, and M.~Yamada, ``{Inflation in random Gaussian
  landscapes},'' \href{http://dx.doi.org/10.1088/1475-7516/2017/05/053}{{\em
  JCAP} {\bfseries 1705} no.~05, (2017) 053},
\href{http://arxiv.org/abs/1612.03960}{{\ttfamily arXiv:1612.03960 [hep-th]}}.

\bibitem{Barrow:1982}
J.~D. {Barrow}, ``{The Isotropy of the Universe},'' {\em Quarterly Journal of
  the Royal Astronomical Society} {\bfseries 23} (Sept., 1982) 344.

\bibitem{Kumekawa:1994gx}
K.~Kumekawa, T.~Moroi, and T.~Yanagida, ``{Flat potential for inflaton with a
  discrete R invariance in supergravity},''
  \href{http://dx.doi.org/10.1143/PTP.92.437, 10.1143/ptp/92.2.437}{{\em Prog.
  Theor. Phys.} {\bfseries 92} (1994) 437--448},
\href{http://arxiv.org/abs/hep-ph/9405337}{{\ttfamily arXiv:hep-ph/9405337
  [hep-ph]}}.

\bibitem{Izawa:1996dv}
K.~I. Izawa and T.~Yanagida, ``{Natural new inflation in broken
  supergravity},'' \href{http://dx.doi.org/10.1016/S0370-2693(96)01638-3}{{\em
  Phys. Lett.} {\bfseries B393} (1997) 331--336},
\href{http://arxiv.org/abs/hep-ph/9608359}{{\ttfamily arXiv:hep-ph/9608359
  [hep-ph]}}.

\bibitem{Harigaya:2013pla}
K.~Harigaya, M.~Ibe, and T.~T. Yanagida, ``{Lower Bound on the Garvitino Mass
  $m_{3/2}>O(100)$ TeV in $R$-Symmetry Breaking New Inflation},''
  \href{http://dx.doi.org/10.1103/PhysRevD.89.055014}{{\em Phys. Rev.}
  {\bfseries D89} no.~5, (2014) 055014},
\href{http://arxiv.org/abs/1311.1898}{{\ttfamily arXiv:1311.1898 [hep-ph]}}.

\bibitem{Agrawal:1997gf}
V.~Agrawal, S.~M. Barr, J.~F. Donoghue, and D.~Seckel, ``{The Anthropic
  principle and the mass scale of the standard model},''
  \href{http://dx.doi.org/10.1103/PhysRevD.57.5480}{{\em Phys. Rev.} {\bfseries
  D57} (1998) 5480--5492},
\href{http://arxiv.org/abs/hep-ph/9707380}{{\ttfamily arXiv:hep-ph/9707380
  [hep-ph]}}.

\bibitem{Hall:2014dfa}
L.~J. Hall, D.~Pinner, and J.~T. Ruderman, ``{The Weak Scale from BBN},''
  \href{http://dx.doi.org/10.1007/JHEP12(2014)134}{{\em JHEP} {\bfseries 12}
  (2014) 134},
\href{http://arxiv.org/abs/1409.0551}{{\ttfamily arXiv:1409.0551 [hep-ph]}}.

\end{thebibliography}\endgroup

\end{document}